%% file: paper.tex
\documentclass[twocolumn]{aa}
\usepackage{graphicx}
\usepackage{natbib}
\usepackage[figuresright]{rotating}
\usepackage{txfonts}
\newcommand{\Deg}{\hbox{${}^\circ$}}
\newcommand{\Min}{\hbox{${}^{\prime}$}}
\newcommand{\Sec}{\hbox{${}^{\prime\prime}$}}
\newcommand{\minp}{\hbox{${}^{\prime}$\llap{.}}}

\newcommand{\hii}{H\,{\sc ii}\rm}
\newcommand{\hei}{He\,{\sc i}\rm}

\newcommand{\siii}{[S\,{\sc iii}]}
\newcommand{\none}{[N\,{\sc i}]}
\newcommand{\nii}{[N\,{\sc ii}]}
\newcommand{\oiii}{[O\,{\sc iii}]}
\newcommand{\oii}{[O\,{\sc ii}]}
\newcommand{\oi}{[O\,{\sc i}]}
\newcommand{\sii}{[S\,{\sc ii}]}
\newcommand{\ariii}{[Ar\,{\sc iii}]}
\newcommand{\neiii}{[Ne\,{\sc iii}]}

\newcommand{\te}{$T_\mathrm{e}$}

\newcommand{\lin}{$\,\lambda$}
\newcommand{\llin}{$\,\lambda\lambda$}
\newcommand{\nodata}{$.\, .\, .$}

\begin{document}
%
%   \title{Chemical abundances and massive stellar populations in 
%   metal-rich extragalactic H\,II regions}

\title{A VLT study of metal-rich extragalactic H\,II regions}

   \subtitle{I. Observations and empirical abundances\thanks{Based on
   observations collected at the European Southern Observatory, Chile,
   proposal ESO 71.B-0236}}

   \author{F. Bresolin
          \inst{1}
\and
D. Schaerer
\inst{2}
          \and
R. M. Gonz\'alez Delgado
\inst{3}
\and
G. Stasi\'nska
\inst{4}
          }

   \offprints{D. Schaerer}

   \institute{Institute for Astronomy, University of Hawaii,
              2680 Woodlawn Drive, Honolulu 96822 USA\\
              \email{bresolin@ifa.hawaii.edu}
         \and
             Observatoire de Geneve, 
51, Ch. des Maillettes, CH-1290 Sauverny, Switzerland\\
             \email{daniel.schaerer@obs.unige.ch}
\and
Instituto de Astrof\'{\i}sica de Andaluc\'{\i}a (CSIC),
Apdo. 3004, 18080 Granada, Spain\\
\email{rosa@iaa.es}
\and
LUTH, Observatoire de Paris-Meudon, 5 Place Jules Jansen,
             92195 Meudon, France\\
\email{grazyna.stasinska@obspm.fr}
             }

   \date{Received / Accepted}

\abstract{We have obtained spectroscopic observations from 3600~\AA\/ to
9200~\AA\/ with FORS at the Very Large Telescope for approximately 70
\hii\/ regions located in the spiral galaxies NGC~1232, NGC~1365,
NGC~2903, NGC~2997 and NGC~5236. These data are part of a project aiming
at measuring the chemical abundances and characterizing the massive
stellar content of metal-rich extragalactic \hii\/ regions. In this
paper we describe our dataset, and present emission line fluxes for the
whole sample. In 32 \hii\/ regions we measure at least one of the
following auroral lines: \sii\lin4072, \nii\lin5755, \siii\lin6312 and
\oii\lin7325. From these we derive electron temperatures, as well as
oxygen, nitrogen and sulphur abundances, using classical empirical
methods (both so-called "\te-based methods" and "strong line methods").
Under the assumption that the temperature gradient does not introduce severe
biases, we find that the most metal-rich nebulae with detected auroral
lines are found at 12\,+\,log(O/H)\,$\simeq$\,8.9, i.e.~about 60\%
larger than the adopted solar value. However, classical abundance determinations
in metal-rich \hii\/ regions may be severely biased and must be tested
with realistic photoionization models. The spectroscopic observations
presented in this paper will serve as a homogeneous and high-quality
database for such purpose.

   }

   \maketitle
%
%________________________________________________________________

\section{Introduction} If the analysis of the emission-line spectra of
extragalactic \hii\/ regions has been essential in the past three
decades to investigate the abundance of heavy elements in star-forming
galaxies, we still lack the observational and theoretical efforts to
adequately understand the metal-rich end (roughly solar and above) of
the nebular abundance scale. While this situation is explained by the
inherent difficulty of measuring abundances in this regime, it is also
true that it affects the study of the inner portions of virtually all
spiral galaxies, in consequence of their radial abundance gradients
(e.g.~\citealt{vila92}, \citealt{zaritsky94}).

The key observational element at low abundance is the strength of the
\oiii\lin4363 auroral line, which allows, in combination with the
nebular \oiii\llin4959,5007 lines, to measure the electron temperature
\te\/ of the gas, upon which the line emissivities strongly depend. It
is well known that, as the cooling efficiency of the gas increases with
the oxygen abundance, the \oiii\/ auroral line becomes too faint to be
observed with the largest telescopes even at modest metallicity. In this
case nebular abundance studies generally rely on {\em statistical
methods}, based on the measurement of strong nebular lines only. The use
of $R_{23}$\,=\,(\oii\lin3727\,+\,\oiii\llin4959,5007)/H$\beta$
(\citealt{pagel79}) has become widespread in this context, however
several different semi-empirical calibrations for this index have been
proposed at high abundance (\citealt{edmunds84}, \citealt{dopita86},
\citealt{mcgaugh91}, \citealt{pilyugin01}, to name just a few).
Additional abundance indicators, which rely on  emission lines present
in the optical spectra of \hii\/ regions other than those from oxygen,
in particular sulphur and nitrogen, have also appeared in the literature
(\citealt{alloin79}, \citealt{diaz00}, \citealt{denicolo02},
\citealt{pettini04}). The usefulness of the statistical methods goes
beyond the derivation of abundance gradients in spirals
(\citealt{pilyugin04}), as these methods find application in chemical
abundance studies of a variety of objects, including low surface
brightness galaxies (\citealt{naray04}) and star-forming galaxies at
intermediate and high redshift, where around-solar oxygen abundances
have been found (\citealt{kobulnicky04}, \citealt{shapley04}).

Recently, starting with the works by \citet{castellanos02} and
\citet{kennicutt03}, and especially with the use of large-aperture
telescopes of the 8m-class by \citet{pindao02}, \citet{garnett04} and
\citet{bresolin04}, it has become possible to measure auroral lines,
such as \nii\lin5755, \siii\lin6312 and \oii\lin7325, at high oxygen 
abundance [up to 12\,+\,log(O/H)\,$\simeq$\,8.9], extending the
application of the {\em direct method} (\te-based) of abundance
determination to the high-metallicity regime, therefore by-passing the
need to use $R_{23}$ or similar indicators for deriving the metallicity
in the inner regions of spirals, as well as allowing empirical
calibrations of the statistical methods at high abundance. These
works conclude that the statistical methods appear to
overestimate abundances around the solar value by as much as 0.2-0.3 dex
[we adopt 12\,+\,log(O/H)$_\odot$\,=\,8.69, following \citealt{allende01}]. There
are, however, uncertainties affecting these \te-based chemical
abundances, that spring from the temperature stratification of
metal-rich \hii\/ regions, which can introduce important biases in the
measured abundances, as shown by \citet{stasinska05}.

In order to resolve some of the pending issues related to metal-rich
extragalactic \hii\/ regions, we started a project in which the first
step is to obtain high-quality spectra of a large sample of these
objects. In this paper we present our observations and analyse them with
classical, empirical methods. Whenever possible, we derive electron
temperatures by using observed auroral lines. We use these temperatures
to obtain direct \te-based abundances for a sizeable sample of \hii\/
regions. We compare these abundances with those derived from statistical
methods based on strong lines only.

In a future paper we will carry out a detailed chemical analysis, with
the aid of photoionization models, for a subset of the sample, in order
to verify the importance of abundance biases at high metallicity and
provide a reliable calibration for strong line methods.

Another paper of this series will deal with the stellar populations
embedded in metal-rich \hii\/ regions. It has been suggested by several
authors that at high metallicity the massive star Initial Mass Function
deviates from the standard Salpeter function, for example with an upper
mass cutoff as low as 30~M$_\odot$ (\citealt{goldader97},
\citealt{bresolin99}, \citealt{thornley00}). However, the presence of
strong wind signatures in the UV spectra of nuclear starbursts is an evidence
against the depletion of massive stars in metal-rich environments
(\citealt{gonzalez02}). Moreover, the detection of Wolf-Rayet (WR) stars
in metal-rich \hii\/ regions allowed \citet{pindao02} to dispute these
claims  (see also \citealt{schaerer00} and \citealt{bresolin02}), and to
show that the progenitors of the WR stars (revealed in the integrated
spectra by their broad emission line features at 4680~\AA\/ and
5808~\AA) are at least as massive as 60~M$_\odot$.

A high-metallicity environment strongly facilitates the formation of WR
stars, through the action of stellar winds driven by radiation scattered
in metal lines. As a consequence, the percentage of \hii\/ regions
expected to display WR features in their spectra varies significantly as
a function of metallicity, from 40\% at 1/5 solar metallicity to 70-80\%
at solar metallicity and above (\citealt{meynet95},
\citealt{schaerer98}). These theoretical predictions are well supported
by recent observations. For example, \citet{crowther04} detected WR
features in nearly 70\% of the $\sim$200 \hii\/ regions they surveyed in
the metal-rich galaxy M83, while 6 out of 10 \hii\/ regions analyzed
spectroscopically in M51 by \citet{bresolin04}, although far from
representing a complete sample, display strong WR emission. Therefore,
investigating metal-rich nebulae, through the properties (flux and
equivalent width) of the emission features of the embedded WR stars and
the statistics of WR stars relative to the total number of ionizing
stars, offers an opportunity to constrain evolutionary models of massive
stars.

In this paper we describe new spectroscopic observations obtained at the
Very Large Telescope of \hii\/ regions in the galaxies NGC~1232,
NGC~1365, NGC~2903, NGC~2997 and NGC~5236 (=\,M83). We present the main
observational data, with tables containing emission line fluxes for
about 70 \hii\/ regions. This paper is structured as follows: we
describe the observations and the data reduction in Section~2, and
discuss the general properties of the \hii\/ regions sample in
Section~3. Electron temperatures are derived from the available auroral
lines in Section~4, and we compute direct abundances of oxygen, nitrogen
and sulphur in Section~5. We summarize our paper in Section~6.

%__________________________________________________________________
\section{Observations and data reduction}

\subsection{Target selection}
For this project we selected galaxies where the available nebular
studies from the literature indicated the presence of high-abundance
\hii\/ regions (\citealt{pagel79}; \citealt{mccall85};
\citealt{vila92}; \citealt{zaritsky94}; \citealt{roy97};
\citealt{vanzee98}; \citealt{bresolin02}).  In most cases, this
judgement has been based on the strength of the oxygen emission lines,
through the use of the semi-empirical abundance indicator $R_{23}$ and
its calibration from different authors (\citealt{edmunds84};
\citealt{dopita86}; \citealt{kobulnicky99};
\citealt{pilyugin01}). Only in the case of NGC~1232 a {\em direct}
measurement of above-solar oxygen abundance in one \hii\/ region was
available, from the detection of the \nii\lin5755 and \siii\lin6312
auroral lines by \cite{castellanos02}. A brief compilation of galactic
parameters for our sample is given in Table~\ref{parameters}.

%_____________________________________________________________
%                                     Table: Galaxy parameters 
%_____________________________________________________________
%
\begin{table*}
\begin{minipage}[t]{\textwidth}
\caption{Galaxy parameters}  
\label{parameters}           
\centering    
\renewcommand{\footnoterule}{}  % to avoid a line before footnotes      
\begin{tabular}{c c c c c c c c}     % 8 columns 
\hline\hline Galaxy & R.A. (2000)\footnote{Source: NASA/IPAC
  Extragalactic Database} & DEC (2000)$^a$ & Morphological$^a$ &
  Distance\footnote{NGC~1232, NGC~2903, NGC~2997: Lyon/Meudon
  Extragalactic Database; NGC~1365: \citet{freedman01}; NGC~5236:
  \citet{thim03}} & $i$\footnote{Source: RC3} & P.A.$^c$ &
  $R_{25}$$^c$ \\
       &             &            & type          & (Mpc) & (deg) &
       (deg) & (arcsec) \\
\hline   
NGC 1232 & 03$^{\rm h}$ 09$^{\rm m}$ 45\fs5 & $-20$\Deg~34\Min~46\Sec & SAB(rs)c & 19.6 & 29 & 108
& 222 \\
NGC 1365 & 03$^{\rm h}$ 33$^{\rm m}$ 36\fs4 & $-36$\Deg~08\Min~25\Sec & SBb(s)b  & 17.2 &
57 & 32 & 337 \\
NGC 2903 & 09$^{\rm h}$ 32$^{\rm m}$ 10\fs1 & $+21$\Deg~30\Min~03\Sec & SB(s)d   & 9.4 &
61 & 17 & 378 \\
NGC 2997 & 09$^{\rm h}$ 45$^{\rm m}$ 38\fs8 & $-31$\Deg~11\Min~28\Sec & SA(s)c   & 9.4 &
41 & 110 & 300 \\
NGC 5236 & 13$^{\rm h}$ 37$^{\rm m}$ 00\fs9 & $-29$\Deg~51\Min~57\Sec & SAB(s)c  & 4.5 &
27 & 45 & 395 \\
\hline                  
\end{tabular}
\end{minipage}
\end{table*}

The \hii\/ regions for the spectroscopic work were selected by examining
narrow-band H$\alpha$ images from various sources, either published or
otherwise available to us. Given the nature of the multi-object
spectroscopy technique adopted for our observations and the presence of
radial abundance gradients in the target galaxies, we have included in
our sample nebulae with different luminosities and chemical abundances,
with those in the central galactic regions likely to approach or exceed
the solar oxygen abundance. When possible, the brightest \hii\/ regions
at a given projected galactocentric distance were chosen, in order to
increase the odds of detecting faint auroral lines and WR stellar features
in emission. $R$-band images obtained at the VLT prior to the
spectroscopic observations were used to measure \hii\/ region positions
and to define the multi-object spectroscopy setups, via the FIMS
software provided by the European Southern Observatory's User Support
Group.

\subsection{Observations} The spectra were obtained in service mode at
the VLT with the FORS2 spectrograph, in the period March to September,
2003. Seeing conditions were typically better than 1.2 arcsec, as
summarized in Table~\ref{observations}. Each galaxy was observed with
one setup composed of up to 19 slitlets, each 1 arcsec wide and 20
arcsec long, distributed over the $6\minp8\times 6\minp8$ FORS field of
view. The complete optical and near-infrared spectra of the \hii\/
regions were obtained with three different grisms: 600B
($\sim3500-5200$\,\AA, 5\,\AA\/ FWHM), 600R ($\sim5000-8500$\,\AA,
6\,\AA\/ FWHM) and 300I ($\sim6500-10000$\,\AA, 11\,\AA\/ FWHM). This
choice of grisms ensured that the auroral and stellar features in the
blue/red part of the spectrum, when detected, were observed with sufficient
spectral resolution for the analysis, while still covering the
near-infrared wavelengths, necessary for measuring the \siii\lin9069
lines. Different total exposure times (divided into two contiguous
exposures) were used for the 5 galaxies, as summarized in
Table~\ref{observations}.

\begin{figure} \centering
\includegraphics[width=0.48\textwidth]{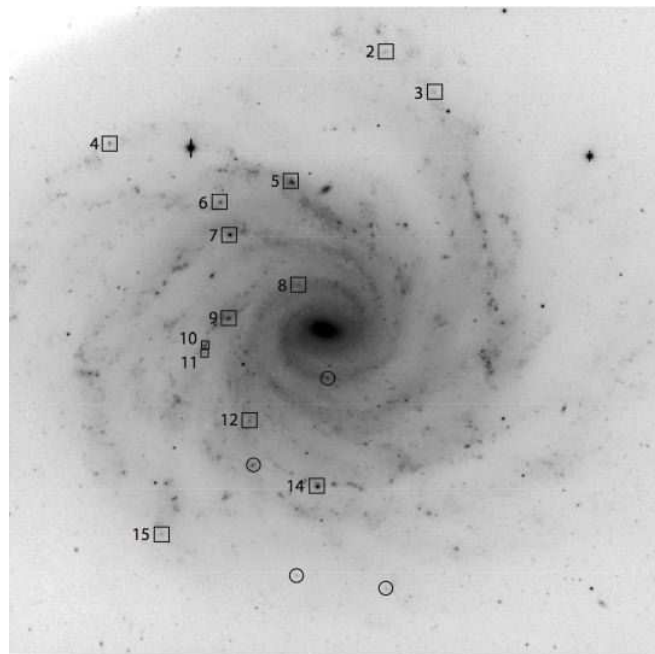} \caption{\hii\/
region identification for NGC~1232. In this and in the following charts,
derived from $R$-band FORS1 or FORS2 images, the slitlet numbers for the
objects marked by squares correspond to those in Tables~\ref{hiiglobal1}
and \ref{hiiglobal2}. The open circles mark additional objects observed
spectroscopically, but not included in the analysis of this paper,
because of the extreme faintness or the absence of emission lines.
Orientation is North to the top and East to the left.  } \label{ngc1232}
\end{figure}

\begin{figure} \centering
   \includegraphics[width=0.48\textwidth]{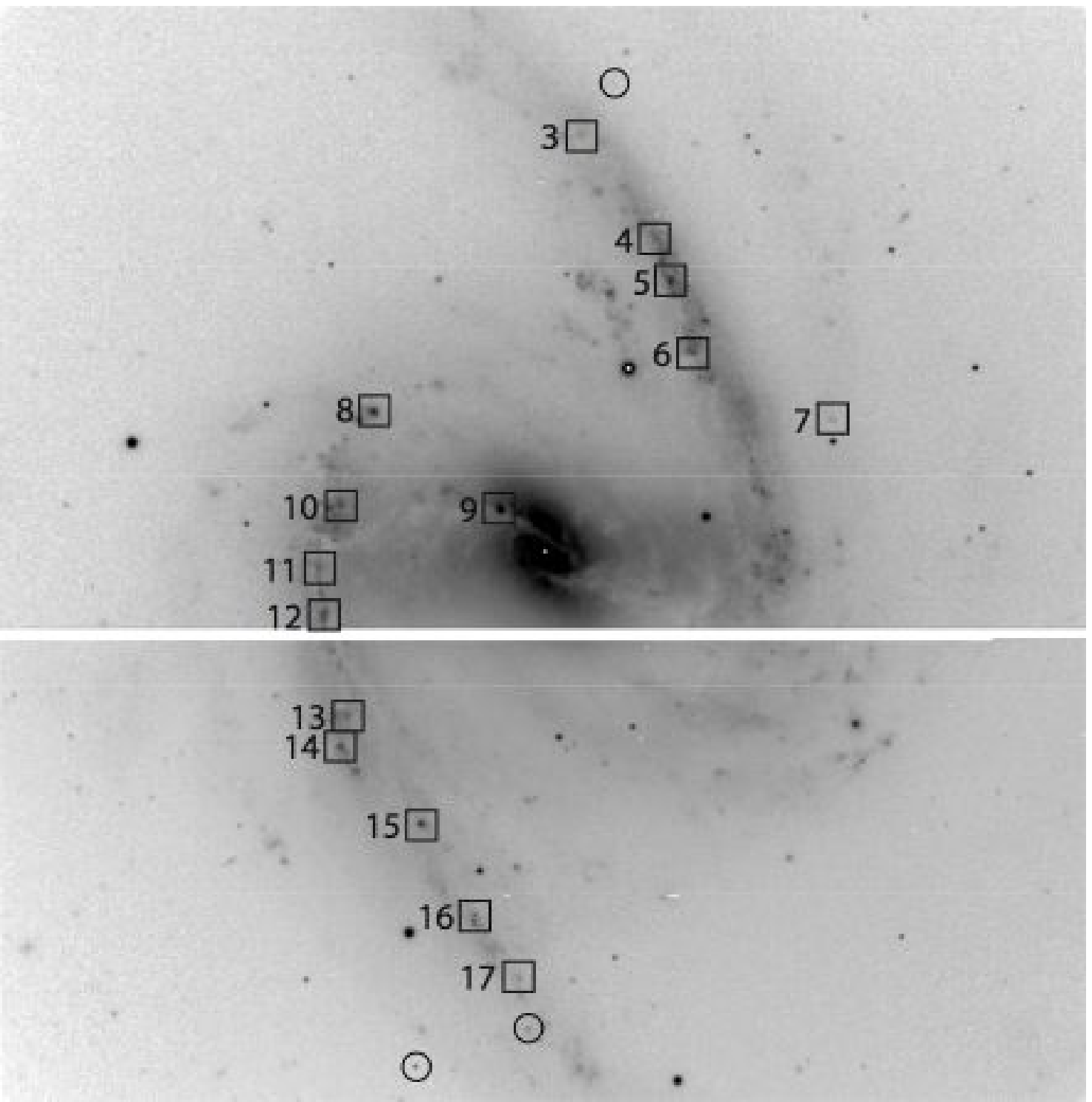} \caption{\hii\/
   region identification for NGC~1365. The gap in the FORS2 CCD mosaic
   runs horizontally.} \label{ngc1365} \end{figure}

\begin{figure}
   \centering
   \includegraphics[width=0.48\textwidth]{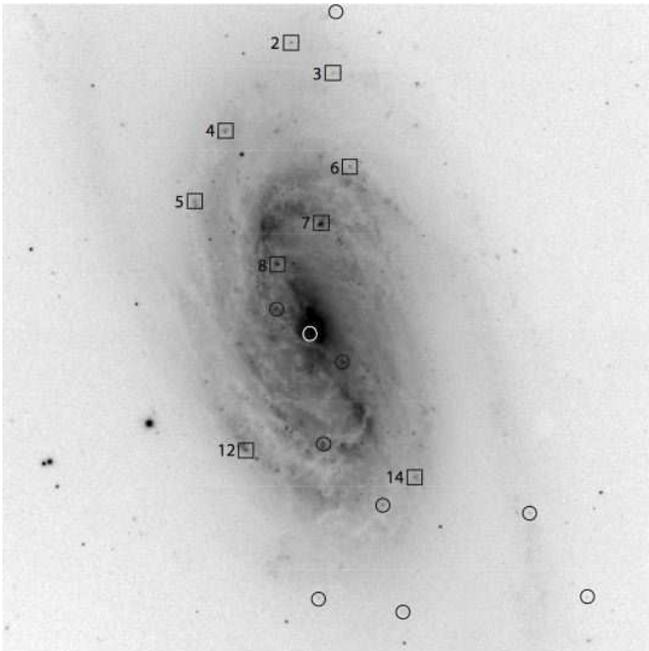}
      \caption{\hii\/ region identification for NGC~2903. }
         \label{ngc2903}
   \end{figure}

\begin{figure} \centering
   \includegraphics[width=0.48\textwidth]{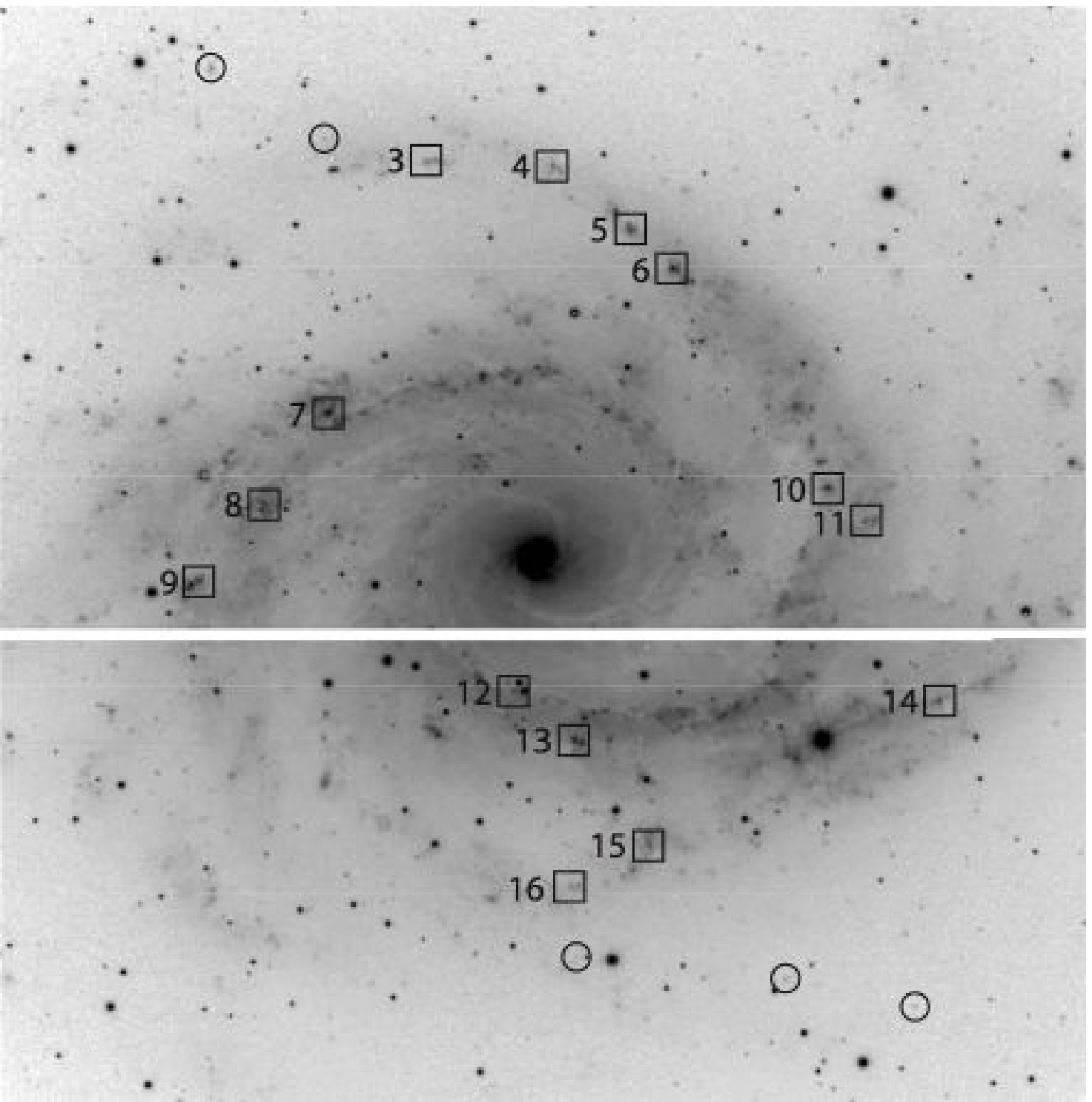} \caption{\hii\/
   region identification for NGC~2997. The gap in the FORS2 CCD mosaic
   runs horizontally.} \label{ngc2997} \end{figure}

\begin{figure}
   \centering
   \includegraphics[width=0.48\textwidth]{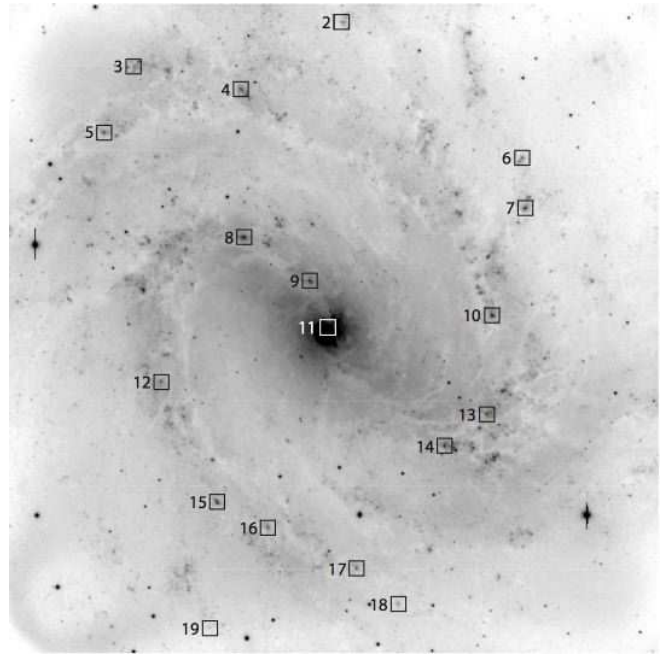}
      \caption{\hii\/ region identification for NGC~5236. }
         \label{ngc5236}
   \end{figure}

Finding charts for the observed \hii\/ regions can be found in
Fig.~\ref{ngc1232}-\ref{ngc5236}, where we have marked with squares the
nebulae analyzed in this paper, and with circles some additional targets
not included in the analysis, due to the  low signal-to-noise of their
spectra, or the heavy contamination by underlying stellar components.
One of these excluded objects is a quasar at redshift $z\simeq2.55$ (see
Appendix~A).

%_____________________________________________________________ 
%                          Figures: HII region finding charts
%-------------------------------------------------------------

%_____________________________________________________________
%

%_____________________________________________________________
%                                          Table: Observations 
%_____________________________________________________________
%
\begin{table*}
\caption{Observing log and sky conditions.}       
\label{observations} 
\centering        
\begin{tabular}{c c c c c c c c c}     % 9 columns 
\hline\hline 
Galaxy & Observing & Sky & \multicolumn{3}{c}{Seeing (arcsec)} &
\multicolumn{3}{c}{Exposure time (sec)}\\
       & date & conditions & 600B & 600R & 300I & 600B & 600R & 300I\\
\hline   
NGC 1232 & Aug 5 2003 & photometric & 0.6--0.8 & 0.6--0.8 & 0.6--0.8 & 1660
& 3060 & 1340 \\

NGC 1365 & Sep 19 2003 & clear & 1.1--1.3 & 0.9--1.1 & 1.2--1.5 & 2000
& 2600 & 1600\\

NGC 2903 & Apr 24 2003 & clear & 0.9--1.0 & 1.5--1.8 & 1.1--1.2 & 800
& 1320 & 680\\

NGC 2997 & May 3,5 2003 & clear, cirrus & 1.0--1.2 & 1.0--1.2 &
1.1--1.2 & 2000 & 2600 & 1600 \\

NGC 5236 & Apr 23 2003 & clear & 0.8--0.9 & 0.9--1.0 & 0.7--0.8 & 650
& 1190 & 560\\

\hline                  
\end{tabular}
\end{table*}
%_____________________________________________________________
%

\subsection{Data reduction} The data reduction was carried out using
standard {\sc iraf}\footnote{{\sc iraf} is distributed by the National
Optical Astronomy Observatories, which are operated by the Association
of Universities for Research in Astronomy, Inc., under cooperative
agreement with the National Science Foundation.} routines, and included
bias and flat field corrections, wavelength and flux calibrations, and
atmospheric extinction correction. The flux calibration provided by the
observed standard star spectra appears satisfactory for the whole
wavelength range, except at the longest wavelengths, above roughly
9200~\AA. The three spectral segments were then flux-normalized using
lines in common: \hei\lin 5876 between 600B and 600R,
H\,$\alpha$+\nii\llin 6548,6583 and \sii\llin 6716,6731 between 600R and
300I. Only in rare instances this procedure introduced corrections
larger than 10\% relative to the scaling provided by the independent
flux calibrations. In those 14 cases where the \hei\lin 5876 was not
included in both the 600B and the 600R spectra, the flux scaling factor
was obtained by requiring that H$\alpha$/H$\beta$\,=\,2.86, as in case B
at \te\, = 10,000~K, after the proper extinction correction, determined
from H$\beta$ and higher-order Balmer lines, had been applied.

For the interstellar extinction correction we used the Balmer decrement
measured by the H$\alpha$, H$\gamma$ and H$\delta$ lines, and the
reddening law of \citet{seaton79}, as parameterized by
\citet{howarth83}, assuming a total-to-selective extinction ratio
$R_\mathrm{V}=A_\mathrm{V}/E_\mathrm{B-V} = 3.1$, and case B theoretical
ratios at 10,000~K (\citealt{hummer87}). We iteratively solved for the
value of c(H$\beta$) and for the absorption originating from the
underlying stellar population, assuming that the equivalent width of the
absorption component is unchanged throughout the Balmer series. The
value for the latter was found to be in the range 0--5\,\AA.  In several
cases the H$\alpha$/H$\beta$ and the H$\delta$/H$\beta$ gave consistent
results, but differing from the extinction measured from
H$\gamma$/H$\beta$. A weighted average for c(H$\beta$) was then adopted.
We also experimented with the reddening law of \citet{cardelli89}, and
found it even more difficult to converge on a value for c(H$\beta$)
using a single value for the absorption equivalent width, although the
estimated extinction was, in general, in fair agreement with that
measured with the Seaton law.

\begin{figure*}
   \centering
   
   \includegraphics[angle=-90,width=1.0\textwidth]{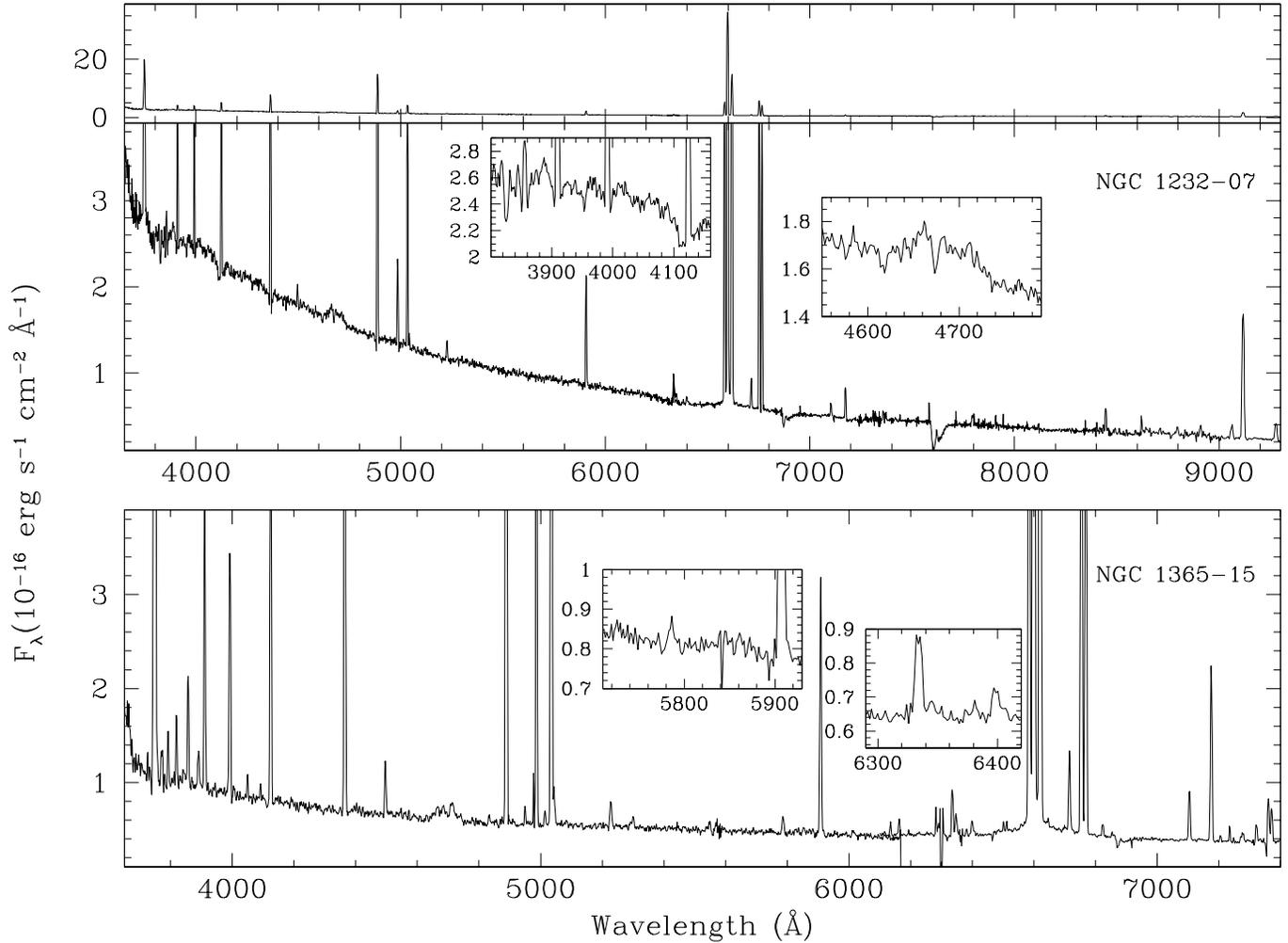}
   \caption{{\em (Top)} The combined spectrum of NGC~1232-07, showing the full extent
   of the spectral coverage of our observations. Two different vertical
   scales are used. The insets show zoomed-in portions of the spectrum,
   where strong stellar features are located: Balmer absorption lines and
   WR emission lines. {\em (Bottom)} Portion of the spectrum observed in
   NGC~1365-15, with the auroral lines \nii\lin 5777 and \siii\lin 6312
   highlighted. } \label{spectra1} \end{figure*}

\begin{figure*}
   \centering

   \includegraphics[angle=-90,width=1.0\textwidth]{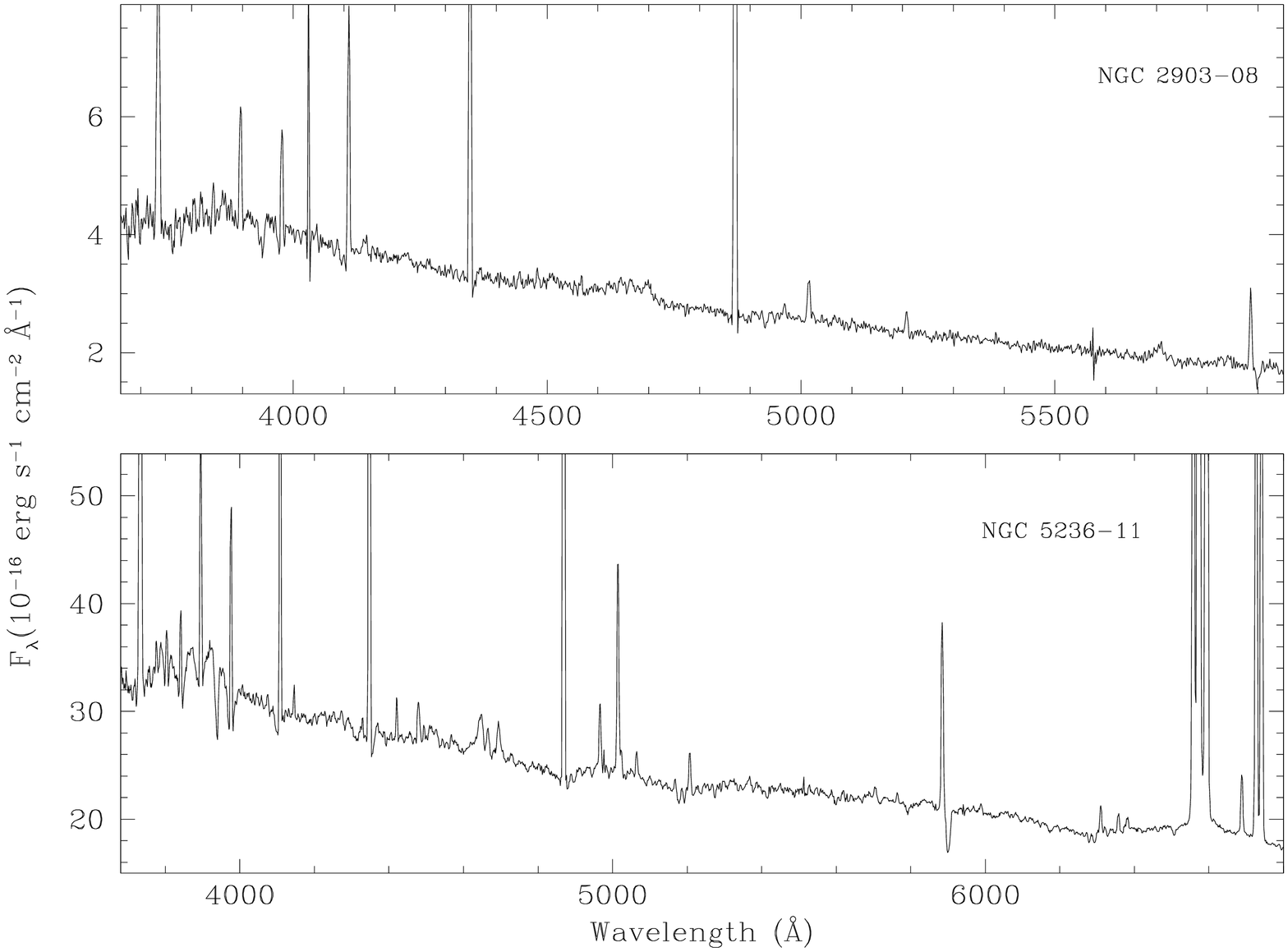}
   \caption{{\em (Top)} The blue portion of the spectrum of NGC~2903-08, a
   low-excitation \hii\/ region, as indicated by the weak
   \oiii\llin4959,5007 emission. Emission features due to WR stars are
   seen at 4680~\AA\/ (WN subtypes), 5696~\AA\/ and 5808~\AA\/ (WC
   subtypes). {\em (Bottom)} Portion of the spectrum of NGC~5236-11, a
   hot-spot \hii\/ region located in the nucleus of the galaxy. Several stellar
   (WR 4680 bump, Balmer absorption lines) and interstellar (Ca\,{\sc
   ii}, Na\,{\sc i}) features are seen, together with auroral lines
   (\nii\lin 5755, \siii\lin 6312). } \label{spectra2} \end{figure*}

We display in Fig.~\ref{spectra1} and \ref{spectra2} a few examples of
\hii\/ region spectra extracted from our sample. The spectrum in the top
panel of Fig.~\ref{spectra1} (NGC~1232-07) shows the complete wavelength
range covered by the combination of the 600B, 600R and 300I grisms.
Zoomed-in examples of stellar features in the blue, namely absorption
components and the WR emission bump, are also included. The bottom panel
shows the blue-red spectral range in NGC~1365-15, where auroral lines
are easily detected: the insets show the \nii\lin 5755 and \siii\lin6312
lines. The top panel of Fig.~\ref{spectra2} shows part of the spectrum
of NGC~2903-08, a low-excitation object (notice the weak
\oiii\llin4959,5007 lines) where WR features are seen at 4686~\AA,
5696~\AA\/ and 5808~\AA. The bottom panel displays the specrum of
NGC~5236-11, a bright hot-spot \hii\/ region in the nucleus of the
galaxy. The WR blue bump, first detected by \citet[their
object A]{bresolin02}, is quite strong. Stellar and interstellar absorption features
are seen throughout this spectrum.

%_____________________________________________________________
%                                 Figure: spectra 1
%-------------------------------------------------------------

%_____________________________________________________________
%

%_____________________________________________________________
%                                 Figure: spectra 2
%-------------------------------------------------------------

%_____________________________________________________________
%

\subsection{Line fluxes: results} In Tables~\ref{hiiglobal1} and
\ref{hiiglobal2} we present for each \hii\/ region in the five target
galaxies the following quantities: slit number (the original slit in the
FORS2 multi-object spectroscopy setup), the offsets from the galaxy
center (in arcsec, measured increasing to the East and North), the
deprojected galactocentric radius (calculated from the observed position
and the galactic parameters of Table~\ref{parameters}), the size of the
spectral extraction window (in arcsec), the extinction c(H$\beta$), the
equivalent width of the nebular H$\beta$ emission and its flux, and the
equivalent width of the Balmer line absorption estimated from the
extinction correction procedure. As mentioned earlier, we do not include
in these two tables those objects which have a S/N ratio that is too low
for a useful analysis, and those objects where, despite a high S/N in
the continuum, the brightest Balmer emission lines were either absent or
almost completely lost in the underlying stellar absorption. This leaves
us with a sample of 69 \hii\/ regions.

Line flux ratios, relative to H$\beta$\,=\,100, for nebular emission
lines of interest are given in Tables~\ref{fluxes1232}-\ref{fluxes5236}.
The associated errors reflect the uncertainties in the flat field
correction and in the flux calibration, as well as the statistical
errors. As these tables show, auroral lines (\sii\lin 4072, \nii\lin
5755, \siii\lin 6312, \oii\lin 7325), which allow the determination of
electron temperatures of the various ions, were measured in 32 \hii\/
regions, nearly half of the whole sample.  The \hei\/ lines  have been
corrected for an average absorption component, following the recipe
given in \citet{kennicutt03}.

\subsection{Line fluxes: comparisons} Several of our target \hii\/
regions have been observed by previous investigators, and we carried out
a survey of the literature, in order to compare our measurements with
the published line fluxes. The quality of the published material is
heterogeneous (in terms of sensitivity, detector, telescope aperture, slit size,
etc.), but a simple comparison can still be useful, in that it could
reveal important systematic effects in the new, deeper observations.  We
have thus extracted measurements of  \oii\lin 3727, \oiii\lin 5007,
\nii\lin 6583 and \sii\llin 6716,6731 from the following papers:

   \[
      \begin{array}{lp{0.75\linewidth}}
         {\rm NGC~1232}  & \citet{vanzee98}\\
         {\rm NGC~1365}  & \citet{pagel79}, \citet{alloin81}, \citet{roy88}, \citet{roy97}\\
         {\rm NGC~2903}  & \citet{mccall85}, \citet{zaritsky94}, \citet{vanzee98}\\
         {\rm NGC~2997}  & \citet{edmunds84}, \citet{walsh89}\\
         {\rm NGC~5236}  & \citet{bresolin02} \\
      \end{array}
   \]

\begin{figure}
   \centering
   \includegraphics[width=0.5\textwidth]{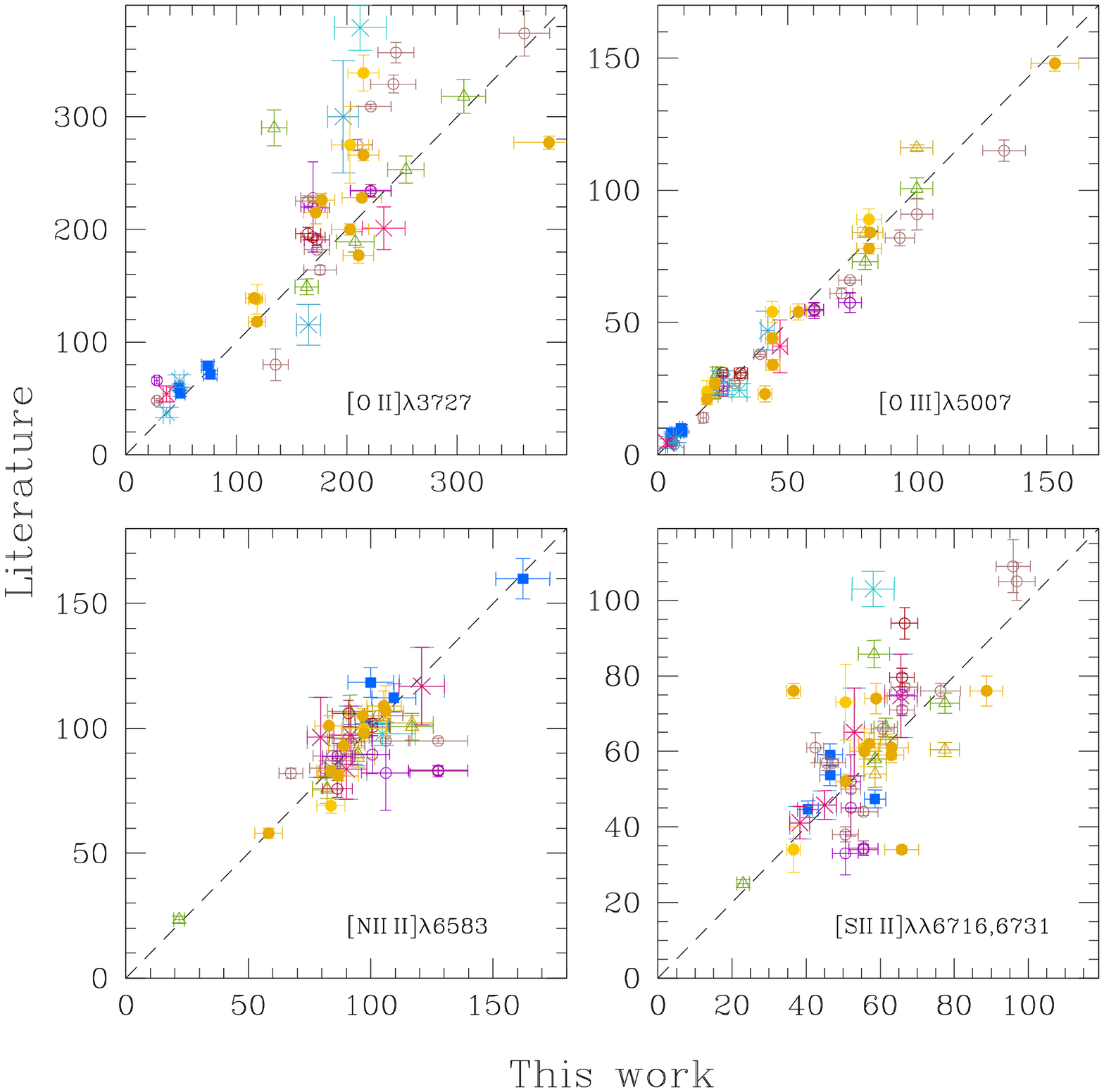}
      \caption{Comparison of reddening-corrrected line intensities (in units of H$\beta$\,=\,100)
      measured in the current work
      ($x$-axis) with published values from the literature
      ($y$-axis). The four panels refer to the strong lines \oii\lin
      3727 {\em (top left)}, \oiii\lin 5007 {\em (top right)},
      \nii\lin 6583 {\em (bottom left)} and \sii\llin 6716,6731 {\em
      (bottom right)}. The five target galaxies shown are: NGC~1232
      (open triangles), NGC~1365 (open circles), NGC~2903 (crosses),
      NGC~2997 (full circles), NGC~5236 (open squares). Different
      colors are used for different comparison data, which are taken
      from the studies mentioned in the text.}
         \label{compare_lines}
   \end{figure}

\noindent The resulting comparison is displayed in
Fig.~\ref{compare_lines}, where the reddening-corrected line intensities
(in units of H$\beta$\,=\,100) from this paper and from the literature
are plotted along the horizontal axis and the vertical axis, respectively. Symbols
with different colors are shown for the different papers used in this
comparison. Excluding for a moment the first panel concerning \oii\lin
3727, we do not find evidence for systematic deviations from the
dashed lines, representing the locations at which the points would lie
in case of a perfect match between our dataset and the published ones. A
similar conclusion could be drawn for the \oii\lin 3727 line comparison,
were it not for a small number of outliers in the top part of the
diagram. Among these are our \hii\/ regions NGC~2903-14 and NGC~1232-10
(compared to \citealt{vanzee98}, which are also discrepant objects in
the panel concerning \sii\llin 6716,6731), NGC~2997-6 (compared
with \citealt{edmunds84}), and a number of objects compared with
\citet{roy97}. While it is difficult to assess the ultimate reason(s)
for these discrepancies, we note that in some cases (e.g.~NGC~2903-14,
NGC~1232-10) there are ambiguities regarding the centering of the slit,
due to multiple, separate bright emission spots. In other cases, there
are likely some problems with the previously published fluxes, as for
the fiber-fed spectrograph observations of \citet{roy97}, as stated by
these authors themselves. Better agreement is, in fact, found with their
imaging spectrophotometry of NGC~1365 (\citealt{roy88}). Finally,
excellent agreement is found with some of the most recent, CCD-based
work used in the comparison (\citealt{bresolin02}, and
\citealt{vanzee98}, once the two problematic objects mentioned above
have been justifiably excluded). Different extinction estimates could
explain some of the discrepancies seen in Fig.~\ref{compare_lines}. The
tighter agreement seen in the \oiii\/ line flux comparison, relative to
the lower-excitation lines, might also be an indication that, at least
in some cases, the effects of varying slit aperture, orientation and
centering can be significant, since higher ionization is produced in
physically smaller nebular volumes, which are more likely to be included
even in narrow slits. The effects of differential atmospheric 
refraction cannot be excluded, either. We simply note that for our new 
\hii\/ region sample such effects are likely to be negligible, thanks to 
the small airmass of the observations ($<1.1$) or the 
approximate alignment of the slits along to the parallactic angle.

%_____________________________________________________________
%                                 Figure: line flux comparison
%-------------------------------------------------------------

%_____________________________________________________________
%

%_____________________________________________________________
%                      Table: HII region global parameters (1) 
%_____________________________________________________________
%
\begin{table*}
\begin{minipage}[t]{\textwidth}
\caption{\hii\/ region global properties: NGC~1232, NGC~1365 and
NGC~2903.} 
\label{hiiglobal1}
\centering \renewcommand{\footnoterule}{} 
\begin{tabular}{c r r r c c c cc }     % 9 columns 
\hline\hline Slit & \multicolumn{2}{c}{Offsets\footnote{In arcsec,
  positive to the East and to the North of the galaxy center.}} &
  $r_0$\footnote{Deprojected galactocentric distance.} & Extraction
  & c(H$\beta$) & \multicolumn{2}{c}{H$\beta$ emission} &
Balmer line
\\ number & EW
  & NS & (arcsec) & width ($''$) &
 & EW (\AA) & Flux\footnote{Measured
  flux. In units of $10^{-16}$
  erg s$^{-1}$ cm$^{-2}$ \AA$^{-1}$.} &
absorption EW (\AA) \\ \hline
\\[-2mm]
\multicolumn{9}{c}{\em NGC~1232}\\[1mm]
\input{ngc1232_global} \\[-2mm]
\multicolumn{9}{c}{\em NGC~1365}\\[1mm]
\input{ngc1365_global} \\[-2mm]
\multicolumn{9}{c}{\em NGC~2903}\\[1mm]
\input{ngc2903_global} \\[-2mm]
\hline
\end{tabular}
\end{minipage}
\end{table*}
%_____________________________________________________________
%

%_____________________________________________________________
%                      Table: HII region global parameters (2) 
%_____________________________________________________________
%
\begin{table*}
\begin{minipage}[t]{\textwidth}
\caption{\hii\/ region global properties: NGC~2997 and
NGC~5236.} 
\label{hiiglobal2}
\centering \renewcommand{\footnoterule}{} 
\begin{tabular}{c r r r c c c cc }     % 9 columns 
\hline\hline Slit & \multicolumn{2}{c}{Offsets\footnote{In arcsec,
  positive to the East and to the North of the galaxy center.}} &
  $r_0$\footnote{Deprojected galactocentric distance.} & Extraction
  & c(H$\beta$) & \multicolumn{2}{c}{H$\beta$ emission} &
Balmer line
\\ number & EW
  & NS & (arcsec) & width ($''$) &
 & EW (\AA) & Flux\footnote{Measured
  flux. In units of $10^{-16}$
  erg s$^{-1}$ cm$^{-2}$ \AA$^{-1}$.} &
absorption EW (\AA) \\ \hline
\\[-2mm]
%\multicolumn{9}{c}{\fbox{\rule[-0.5mm]{0cm}{3mm}NGC~2997}}\smallskip\\ 
\multicolumn{9}{c}{\em NGC~2997}\\[1mm]
\input{ngc2997_global} \\[-2mm]
\multicolumn{9}{c}{\em NGC~5236}\\[1mm]
\input{ngc5236_global} \\[-2mm]
\hline
\end{tabular}
\end{minipage}
\end{table*}
%_____________________________________________________________
%

%%%%%%%%%%%%%%%%%%%%%%%%%%%%%%%%%%
%_____________________________________________________________
%                                                                                  Table: HII region fluxes
%_____________________________________________________________
%
% run nodata.sh to substitute 0.0 with /nodata
%
\begin{sidewaystable*}
\caption{Dereddened nebular emission line fluxes: NGC~1232.} 
\label{fluxes1232}
\centering \renewcommand{\footnoterule}{} 
\begin{tabular}{c c c c c c c c c c c c}     % 11 columns 
\hline\hline 
Object & \oii\lin 3727 & \neiii\lin3869 &
 H$\delta$ & \sii\lin\ 4072 & 
H$\gamma$ & \hei\lin 4471 & 
\oiii\lin 4959 & \oiii\lin5007 & 
\none\lin 5200 & \nii\lin 5755 &
 \hei\lin 5876 \\
\hline
\input{ngc1232_b.tex}
\hline\\
 & \oi\lin 6300 & \siii\lin 6312 & \nii\lin 6548 &
 H$\alpha$ & \nii\lin6583 &
 \hei\lin 6678 & \sii\lin 6716 &
 \sii\lin 6731 & \ariii\lin 7135 &
 \oii\lin 7325 & \siii\lin 9069 \\
\hline
\input{ngc1232_r.tex}
\hline
\end{tabular}
%\begin{flushleft}
%NOTES: \#11 is a second object in slitlet 10
%\end{flushleft}
\end{sidewaystable*}
%_____________________________________________________________
%
\begin{sidewaystable*}
\caption{Dereddened nebular emission line fluxes: NGC~1365.} 
\label{fluxes1365}
\centering \renewcommand{\footnoterule}{} 
\begin{tabular}{c c c c c c c c c c c c}     % 11 columns 
\hline\hline 
Object & \oii\lin 3727 & \neiii\lin3869 &
 H$\delta$ & \sii\lin\ 4072 & 
H$\gamma$ & \hei\lin 4471 & 
\oiii\lin 4959 & \oiii\lin5007 & 
\none\lin 5200 & \nii\lin 5755 &
 \hei\lin 5876 \\
\hline
\input{ngc1365_b.tex}
\hline\\
 & \oi\lin 6300 & \siii\lin 6312 & \nii\lin 6548 &
 H$\alpha$ & \nii\lin6583 &
 \hei\lin 6678 & \sii\lin 6716 &
 \sii\lin 6731 & \ariii\lin 7135 &
 \oii\lin 7325 & \siii\lin 9069 \\
\hline
\input{ngc1365_r.tex}
\hline
\end{tabular}
\end{sidewaystable*}
%_____________________________________________________________
%
\begin{sidewaystable*}
\caption{Dereddened nebular emission line fluxes: NGC~2903.} 
\label{fluxes2903}
\centering \renewcommand{\footnoterule}{} 
\begin{tabular}{c c c c c c c c c c c c}     % 11 columns 
\hline\hline 
Object & \oii\lin 3727 & \neiii\lin3869 &
 H$\delta$ & \sii\lin\ 4072 & 
H$\gamma$ & \hei\lin 4471 & 
\oiii\lin 4959 & \oiii\lin5007 & 
\none\lin 5200 & \nii\lin 5755 &
 \hei\lin 5876 \\
\hline
\input{ngc2903_b.tex}
\hline\\
 & \oi\lin 6300 & \siii\lin 6312 & \nii\lin 6548 &
 H$\alpha$ & \nii\lin6583 &
 \hei\lin 6678 & \sii\lin 6716 &
 \sii\lin 6731 & \ariii\lin 7135 &
 \oii\lin 7325 & \siii\lin 9069 \\
\hline
\input{ngc2903_r.tex}
\hline
\end{tabular}
\end{sidewaystable*}
%_____________________________________________________________
%
\begin{sidewaystable*}
\caption{Dereddened nebular emission line fluxes: NGC~2997.} 
\label{fluxes2997}
\centering \renewcommand{\footnoterule}{} 
\begin{tabular}{c c c c c c c c c c c c}     % 11 columns 
\hline\hline 
Object & \oii\lin 3727 & \neiii\lin3869 &
 H$\delta$ & \sii\lin\ 4072 & 
H$\gamma$ & \hei\lin 4471 & 
\oiii\lin 4959 & \oiii\lin5007 & 
\none\lin 5200 & \nii\lin 5755 &
 \hei\lin 5876 \\
\hline
\input{ngc2997_b.tex}
\hline\\
 & \oi\lin 6300 & \siii\lin 6312 & \nii\lin 6548 &
 H$\alpha$ & \nii\lin6583 &
 \hei\lin 6678 & \sii\lin 6716 &
 \sii\lin 6731 & \ariii\lin 7135 &
 \oii\lin 7325 & \siii\lin 9069 \\
\hline
\input{ngc2997_r.tex}
\hline
\end{tabular}
\end{sidewaystable*}
%_____________________________________________________________
%
\begin{sidewaystable*}
\caption{Dereddened nebular emission line fluxes: NGC~5236.} 
\label{fluxes5236}
\centering \renewcommand{\footnoterule}{} 
\begin{tabular}{c c c c c c c c c c c c}     % 11 columns 
\hline\hline 
Object & \oii\lin 3727 & \neiii\lin3869 &
 H$\delta$ & \sii\lin\ 4072 & 
H$\gamma$ & \hei\lin 4471 & 
\oiii\lin 4959 & \oiii\lin5007 & 
\none\lin 5200 & \nii\lin 5755 &
 \hei\lin 5876 \\
\hline
\input{ngc5236_b.tex}
\hline\\
 & \oi\lin 6300 & \siii\lin 6312 & \nii\lin 6548 &
 H$\alpha$ & \nii\lin6583 &
 \hei\lin 6678 & \sii\lin 6716 &
 \sii\lin 6731 & \ariii\lin 7135 &
 \oii\lin 7325 & \siii\lin 9069 \\
\hline
\input{ngc5236_r.tex}
\hline
\end{tabular}
\begin{flushleft}
NOTES: NGC~5236-05: \oii\lin 3727 is a lower limit (line at CCD's edge). NGC~5236-13: only 600B spectrum available.
\end{flushleft}
\end{sidewaystable*}
%_____________________________________________________________
%

%__________________________________________________________________
\section{Empirical diagrams}

\subsection{General properties of the nebulae}

\begin{figure} \centering
   \includegraphics[width=0.47\textwidth]{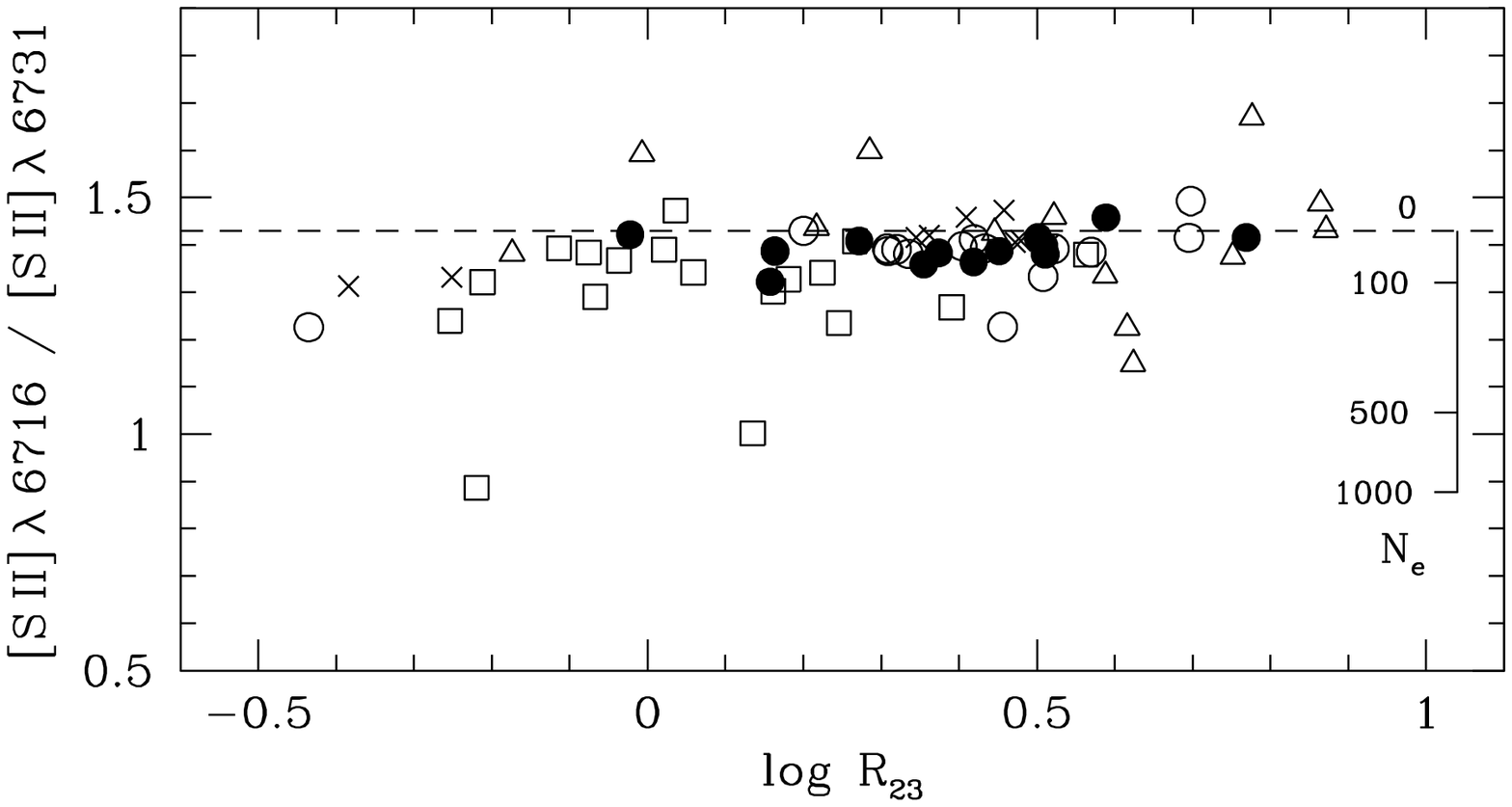} \caption{The electron
   density-sensitive ratio \sii\lin6716/\sii\lin6731 plotted against the
   empirical abundance indicator $R_{23}$ for our \hii\/ region sample.
   The scale on the right provides an approximate density scale (in
   cm$^{-3}$). The zero-density limit is indicated by the dashed line.
   Here and in the following diagrams we use the following symbols to
   differentiate nebulae in the different  galaxies: NGC~1232 {\em (open
   triangles)}, NGC~1365 {\em (open circles)}, NGC~2903 {\em (crosses)},
   NGC~2997 {\em (filled circles)} and NGC~5236 {\em (open squares)}. }
   \label{density} \end{figure}

We can quickly assess some general properties of the \hii\/ region
sample and the quality of the data by looking at diagrams
involving a number of crucial line ratios. In Fig.~\ref{density} we show
the density-sensitive ratio \sii\lin 6716/\sii\lin 6731 as a function of
the abundance-sensitive indicator $R_{23}$. The sulphur line ratio
reaches a 'zero-density limit' at \sii\lin 6716/\sii\lin 6731=1.43
(\te=10,000~K), shown by the dashed line. Almost all of the observed
nebulae lie at this limit or just slightly below, with corresponding
electron densities up to a few hundred particles\,cm$^{-3}$ (as shown by
the density scale on the right). The highest densities are encountered
for two objects in NGC~5236: the central hot-spot \hii\/ region \#11
($N_\mathrm{e}\simeq 1000$ cm$^{-3}$) and the inner-disk \hii\/ region
\#13. The results displayed in this diagram justify the low-density
assumption made for the subsequent analysis of the \hii\/ region sample.

\begin{figure} \centering
   \includegraphics[width=0.47\textwidth]{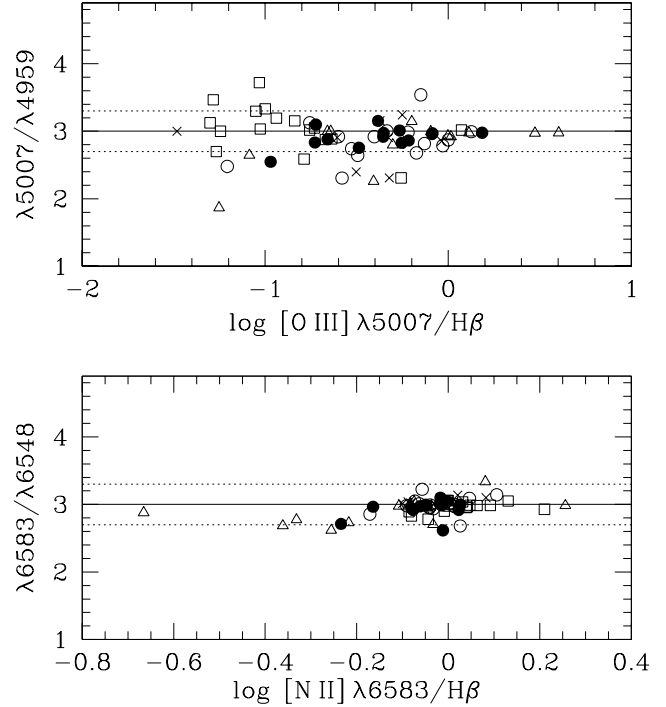} \caption{The ratio
   between the measured \oiii\lin5007 and \oiii\lin4959 fluxes {\em
   (top)} and between the \nii\lin6583 and \nii\lin6548 fluxes {\em
   (bottom)}, compared with the theoretical expectation
   (\oiii\lin5007/\oiii\lin4959 = \nii\lin6583/\nii\lin6548\,=\,3),
   drawn as a continuous line. The dotted lines show  the $\pm$10\%
   deviations from the predicted value. } \label{oiii} \end{figure}

According to the relative radiative transition probabilities in the
O$^{2+}$ and N$^+$ ions, we expect that the line ratios
\oiii\lin5007/\oiii\lin4959 and \nii\lin6583/\nii\lin6548 be nearly
equal to 3. Fig.~\ref{oiii} shows that this is indeed the case. The
dot-dashed lines show the $\pm$10\% deviation from the predicted value.
The higher dispersion in the \oiii\/ doublet line ratio can be explained
by the fact that these lines are generally fainter than the \nii\/
lines.

\begin{figure} \centering
   \includegraphics[width=0.47\textwidth]{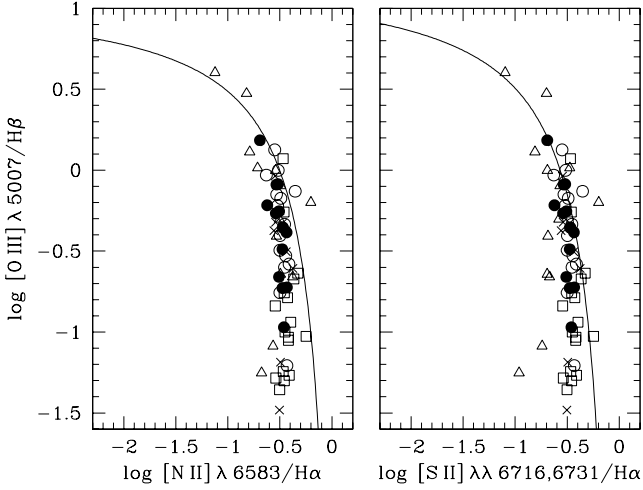} \caption{Nebular
   diagnostic diagrams showing the excitation sequence of our sample. As
   a function of $\log$(\oiii\lin5007/H$\beta$) we plot
   $\log$(\nii\lin6583/H$\alpha$) {\em (left)} and
   $\log$(\sii\llin6716,6731/H$\alpha$) {\em (right)}. The curves
   represent the theoretical upper boundaries calculated by
   \citet{dopita00}. } \label{baldwin} \end{figure}

The excitation properties of the \hii\/ regions are summarized in the
diagrams shown in Fig.~\ref{baldwin}, where the line ratios \nii\lin
6583/H$\alpha$ and \sii\llin6716,6731/H$\alpha$, both involving low
excitation metal lines, are plotted against  \oiii\lin5007/H$\beta$. The
\hii\/ region sequence is extremely tight in both cases, and comprises
objects of mostly low excitation, as expected from the selection of the
targets. High-excitation objects ($\log$\oiii\lin5007/H$\beta$\,$>$\,0)
would, in fact, populate the upper part of the diagram (see similar
plots in \citealt{bresolin02} and \citealt{kennicutt00}), where the
theoretical upper boundaries  from \citet[shown here by full
lines]{dopita00}  turn sharply to the left.

\begin{figure} \centering
   \includegraphics[width=0.47\textwidth]{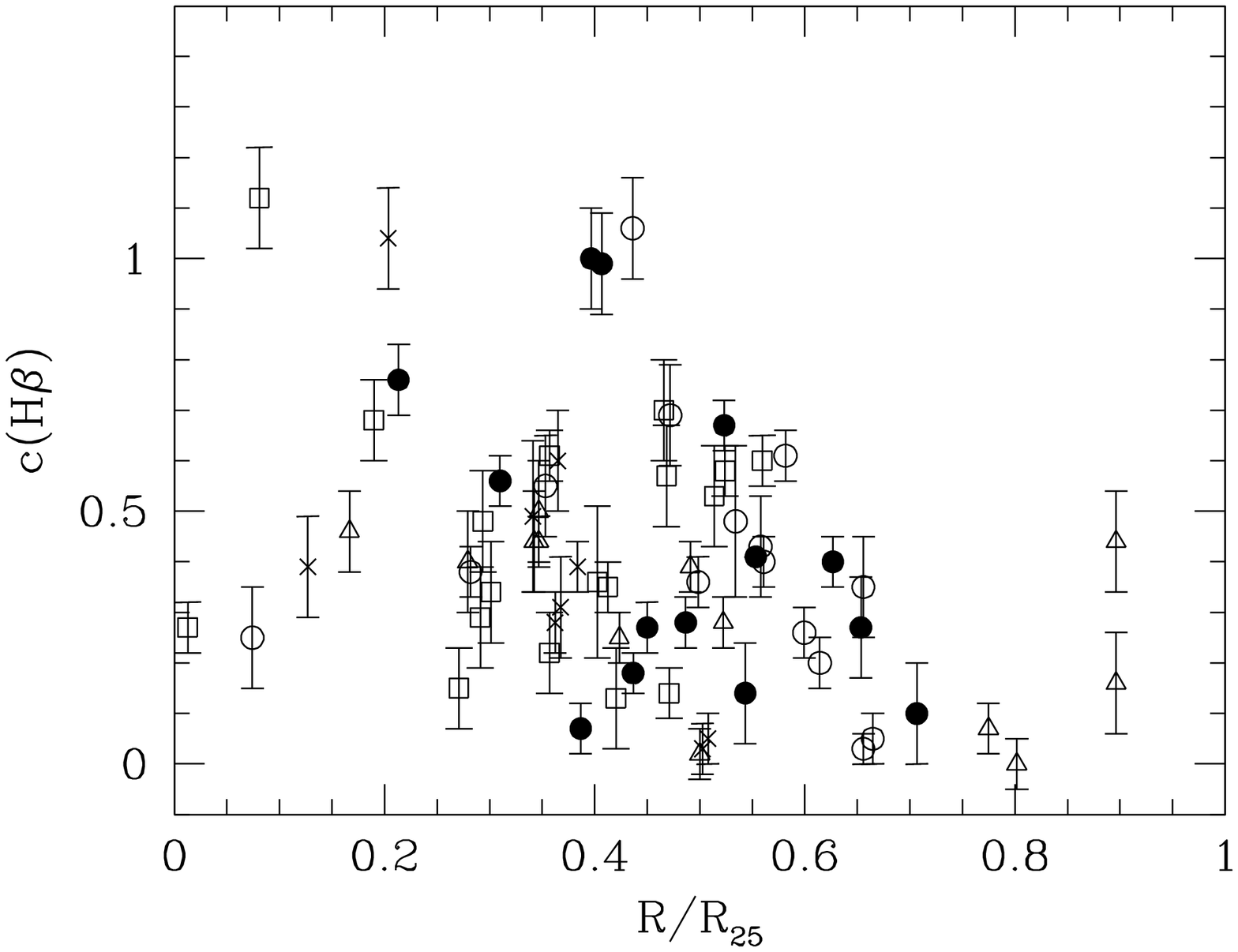} \caption{The radial
   distribution of the extinction c(H$\beta$) within the target
   galaxies. The deprojected radial distances of the individual \hii\/
   regions have been normalized to the isophotal radius $R_{25}$ of the
   parent galaxy. } \label{c} \end{figure}

The nebular extinction c(H$\beta$) appears to be in the typical range
observed in extragalactic \hii\/ regions. Its radial distribution within
the five galaxies is shown in Fig.~\ref{c}, using the galactocentric
distance normalized to the galactic isophotal radius. There is a slight
tendency for larger values of the extinction towards the central regions
of the galaxies, at least in the sense that objects with very low
c(H$\beta$) are found only at $R/R_{25}>0.4$.

%_____________________________________________________________
%                                 Figure: electron density
%-------------------------------------------------------------
   
%_____________________________________________________________
%

%_____________________________________________________________
%                                 Figure: OIII & NII ratios
%-------------------------------------------------------------

%_____________________________________________________________
%

%_____________________________________________________________
%                                 Figure: diagnostic diagrams
%-------------------------------------------------------------

%_____________________________________________________________
%

%_____________________________________________________________
%                                 Figure: c(Hbeta) vs radius
%-------------------------------------------------------------

%_____________________________________________________________
%

\subsection{Abundance estimates from statistical methods} Abundance
estimates of \hii\/ regions can be made by statistical methods based on
strong lines and by direct methods based on the measurement of the
electron temperature. Until recently, the latter methods could not be
applied for metal-rich \hii\/ regions, because the lines necessary to
derive the electron temperature were too weak to be measured. There is
however a growing amount of data that allow such measurements in regions
with very faint auroral lines (\citealt{bresolin04},
\citealt{kennicutt03}, \citealt{pindao02}). Our VLT spectra allow this
for a number of objects. In the following, we use our data to derive
abundances with the methods described by \citet{bresolin04}. However, as
noted by \citet{stasinska05}, these methods are likely to produce strong
biases in metal-rich \hii\/ regions, and we postpone any strong
astrophysical implication of our results to a future paper, where we
will discuss in detail the elemental abundances in our set of objects.
For illustrative purposes, we first derive abundances using published
strong line calibrations, but keeping in mind that such calibrations are
extremely uncertain at the high-abundance end.

Out of the different statistical methods found in the literature, we
considered the following: $R_{23}$\,=\,(\oii\lin
3727\,+\,\oiii\llin4959,5007)/H$\beta$ (\citealt{pagel79}),
$S_{23}$\,=\,(\sii\llin 6716,6731\,+\,\siii\llin9069,9532)/H$\beta$ 
(\citealt{diaz00}), $N2$\,=\,log\,(\nii\lin6583/H$\alpha$)
(\citealt{denicolo02}) and
$O3N2$\,=\,log\,\{(\oiii\lin5007/H$\beta$)/(\nii\lin6583/H$\alpha$)\}
(\citealt{alloin79}, \citealt{pettini04}). For $S_{23}$, since we lacked
the sulphur \lin9532\/ line measurements, we estimated the intensity of
this line from \lin9069 and the theoretical ratio
\lin9532/\lin9069=2.44.

\begin{figure}
   \centering

   \includegraphics[width=0.47\textwidth]{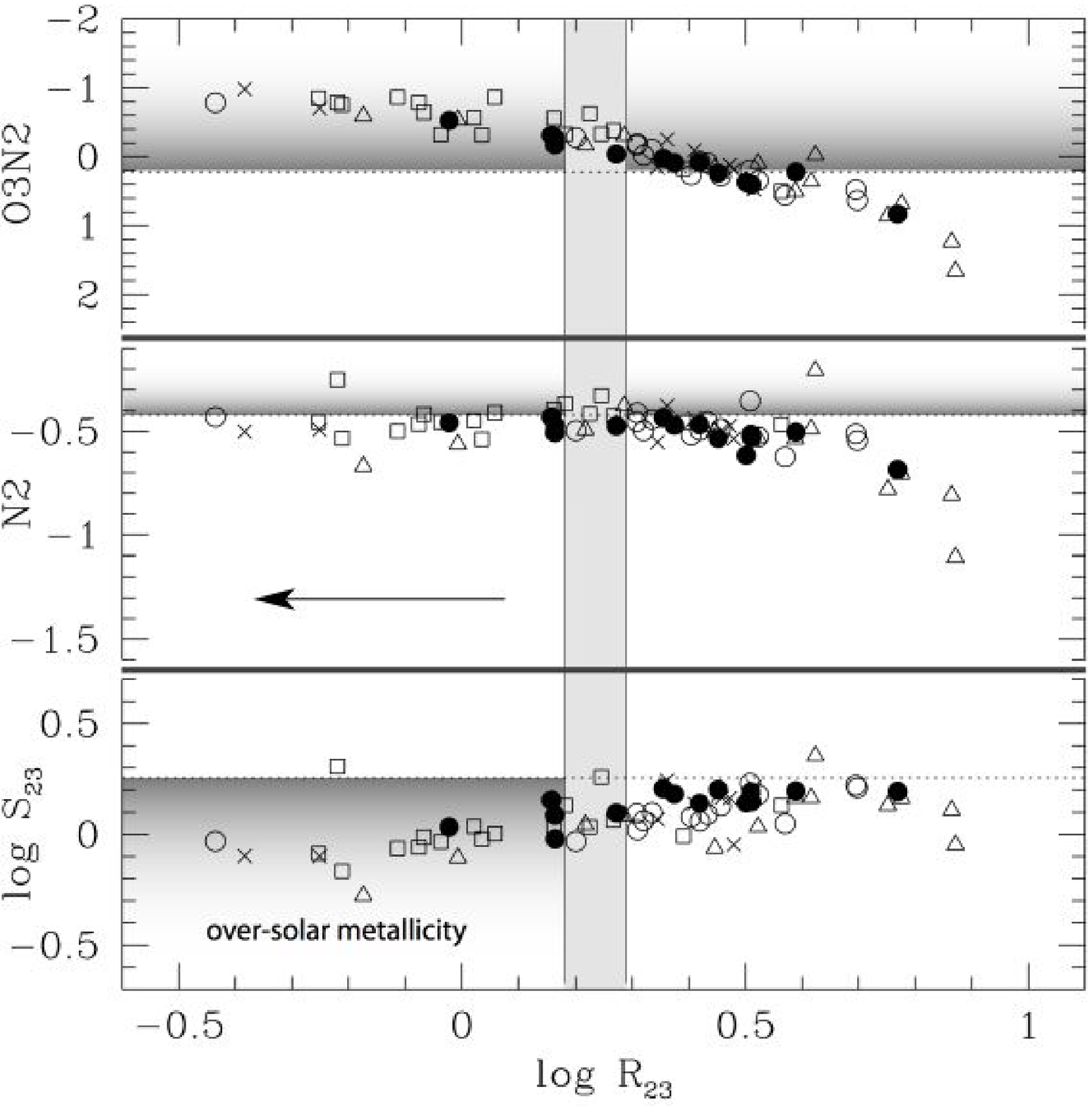}
   \caption{Comparison of statistical abundance indicators: $R_{23}$
   plotted against $O3N2$ {\em(top)}, $N2$ {\em (middle)} and
   log\,$S_{23}$ {\em (bottom)}. The horizontal dotted lines show the
   index value corresponding to the solar O/H abundance
   [12\,+\,log(O/H)$_\odot$\,=\,8.69, \citealt{allende01}], according to
   the calibrations of \citet[$O3N2$, $N2$]{pettini04} and
   \citet[$S_{23}$]{diaz00}. The shaded areas below or above these lines
   define the regions of over-solar metallicity. The vertical light-grey
   band represents the solar O/H value derived from the $R_{23}$
   calibration of \citet{pilyugin01}, for the range of the excitation
   parameter ($P$\,=\,0.1--0.3) which  comprises  the majority of the
   \hii\/ regions in our sample. The arrow shows the direction of
   increasing oxygen abundance according to the $R_{23}$ method. }
   \label{statistical} \end{figure}

The relationship among these different abundance indicators is shown in
Fig.~\ref{statistical}, where we have chosen to plot $R_{23}$ against
the remaining indicators. The dotted lines provide the values
corresponding to the solar abundance, 12\,+\,log(O/H)$_\odot$\,=\,8.69
(\citealt{allende01}), when using the calibrations of the different
indexes from \citet[$O3N2$ and $N2$]{pettini04} and
\citet[$S_{23}$]{diaz00}. It should be noted that the latter indicator
is, like $R_{23}$, non-monotonic, so that a decrease of $S_{23}$ below
log\,$R_{23}=0.3$ (roughly corresponding to the solar O/H value,
according to the \citealt{pilyugin01} calibration) corresponds to an
increase in the oxygen abundance (see \citealt{diaz00}). We also point
out that virtually all of the \hii\/ regions analyzed here belong to the
upper branch of $R_{23}$, following the condition
\nii\lin6583/\oii\lin3727\,$>$\,0.1 to define upper-branch objects
(\citealt{vanzee98}).
 
The diagrams in Fig.~\ref{statistical} suggest that our \hii\/ region
sample contains a good number of high abundance objects, although
the well-known uncertainties in the calibration of the strong line
methods, especially at the metal-rich end, prevent us from providing an
accurate metallicity scale. For example, both $O3N2$ and $S_{23}$ would
indicate the presence of many \hii\/ regions with oxygen abundance well
over the solar value, while $N2$ seems to level off at the solar value
for the majority of the sample.

\begin{figure}
   \centering

   \includegraphics[width=0.47\textwidth]{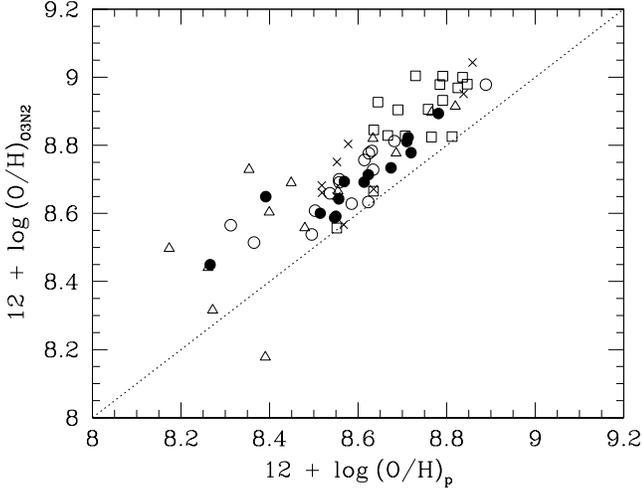} \caption{Oxygen
   abundance from statistical methods: the $P$-method
   (\citealt{pilyugin01}) against $O3N2$ (\citealt{pettini04}). }
   \label{pilyugin} \end{figure}

\begin{figure}
   \centering

   \includegraphics[width=0.47\textwidth]{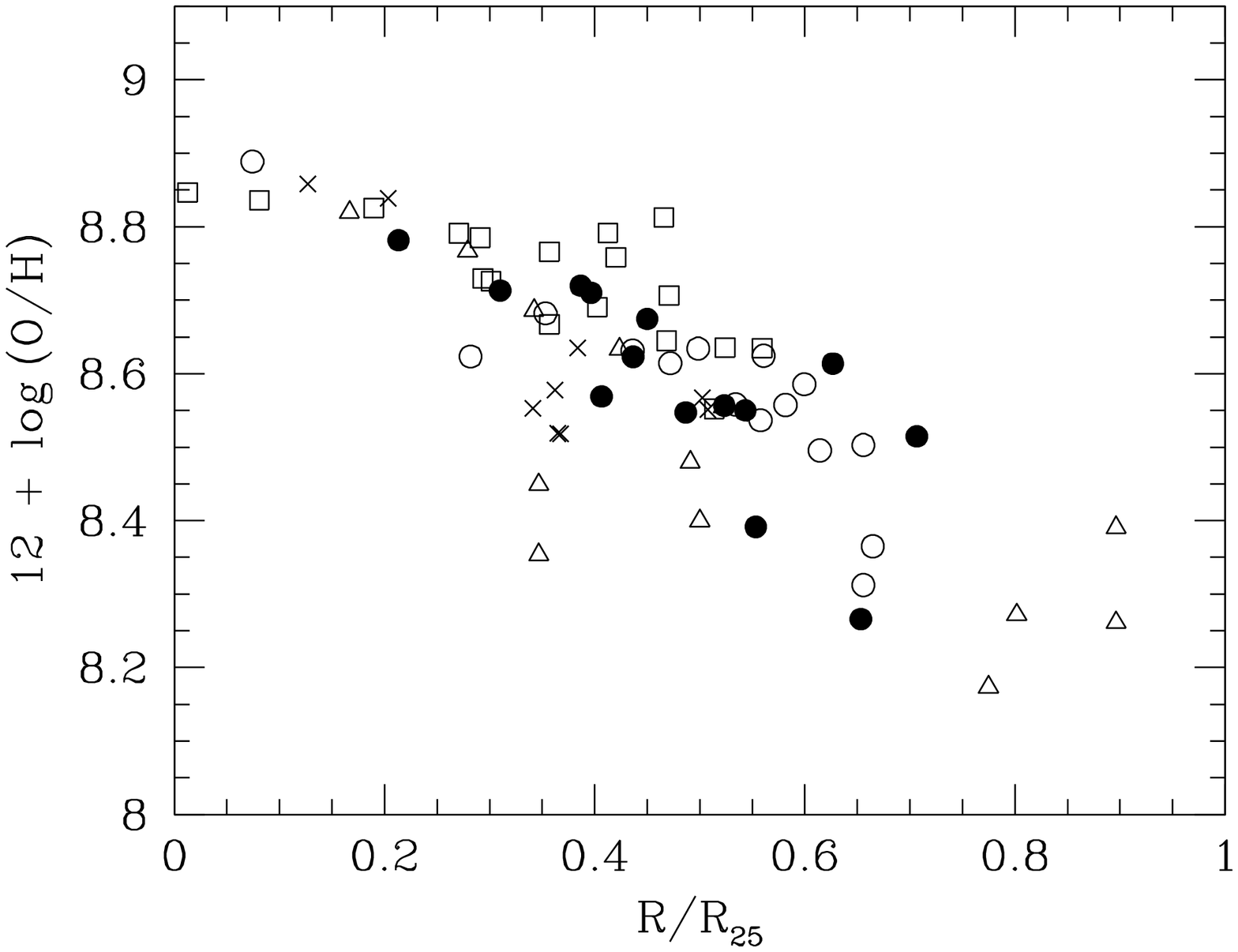} \caption{The
   radial oxygen abundance gradients in the 5 galaxies, estimated via
   the $P$-method of \citet{pilyugin01}. The deprojected radial
   distances of the \hii\/ regions are normalized to the isophotal
   radius of the parent galaxy. } \label{radial} \end{figure}

In order to quantify the oxygen abundances from empirical methods, we
considered the $R_{23}$ indicator, as calibrated by \citet{pilyugin01},
and $O3N2$, as calibrated by \citet{pettini04}. In the former case, we
adopted the upper branch (high metallicity) version of the calibration,
which is applicable when the estimated abundance is 12 +
log(O/H)\,$>$\,8.2 (true for all objects in the sample, except for
NGC~1232-15). The comparison between the oxygen abundances obtained from
the two indicators is displayed in Fig.~\ref{pilyugin}. An offset of
approximately 0.1 dex between the two methods  is immediately apparent.
According to this diagram, the most metal-rich \hii\/ regions in our
sample have an abundance of 12\,+\,log(O/H)\,$\simeq$\,8.9-9.0, which is
approximately twice the currently accepted solar value. Finally, we
display in Fig.~\ref{radial} the radial oxygen abundance gradients for
the target galaxies, as estimated from Pilyugin's $P$-method.
Qualitatively these gradients appear quite similar to each other, even
though differences in the slopes can be found: note, for example, the
somewhat flatter gradient in NGC~5236 (open squares)  compared to the
remaining galaxies.

%_____________________________________________________________
%                                                             Figure: R23 and others
%-------------------------------------------------------------

%_____________________________________________________________
%

%_____________________________________________________________
%                                                             Figure: Pilyugin vs O3N2
%-------------------------------------------------------------

%_____________________________________________________________
%

%_____________________________________________________________
%                                                             Figure: Pilyugin radial
%-------------------------------------------------------------

%_____________________________________________________________
%

\section{Auroral lines and electron temperatures} In this section we
apply the standard technique of measuring nebular electron temperatures
from line ratios involving auroral lines. As seen in
Tables~\ref{fluxes1232}-\ref{fluxes5236}, we have measured one or more
of the  \oii\lin7325, \nii\lin5755, \siii\lin6312 and \sii\lin4072 lines
in several \hii\/ regions from our sample. These can be combined with
stronger lines to form different  line ratios, in short \oii\/3727/7325,
\nii\/(6548,6584)/5755, \siii\/(9069,9532)/6312 and
\sii\/(6716,6731)/4072, which are highly sensitive to the electron
temperature. Once the temperature-sensitive emissivities are calculated
from \te, the abundance of the various chemical elements can be derived.
At high metallicity, however, the auroral lines do not necessarily
provide a good measure of \te, due to the biases introduced by the
presence of temperature gradients within the nebulae
(\citealt{stasinska05}). The complications that these effects introduce
on the derivation of chemical abundances will be treated in a separate
paper. Here we will follow the standard procedure, as if these biases
were not present, but with the warning that the electron temperatures
and abundances derived can be erroneous. Ultimately, we will need to
verify our direct abundances, as derived here, with detailed nebular
models.

Electron temperatures have been obtained from the line ratios listed
above using the five-level atom program {\em nebular} in {\sc
iraf/stsdas}  v.\,3.1 (\citealt{shaw95}). The atomic data adopted are
those included in the May 1997 version of {\em nebular}, except for the
update of the S\,{\sc iii} collisional strengths from \citet{tayal99}.
Electron temperatures were obtained from as many lines as possible for
32 \hii\/ regions, where at least one auroral line was detected. These
temperatures are listed in Table~\ref{tetable}, where we prefer to use
T(7325) instead of T\oii, and similarly for the other lines, to indicate
the possibility that these temperatures might be different from the real
ionic temperatures. The \oii\lin7325 line is usually the strongest
auroral line in the measured spectra, and was detected for all \hii\/
regions included in Table~\ref{tetable}. On the other hand,
\sii\lin4072, from which T(4072) was derived, has been seldom detected,
its measurement made difficult by low signal-to-noise in the spectra.
Both T(5755) and T(6312) were computed for about half of the sample in
Table~\ref{tetable}.

\begin{figure}
   \centering

   \includegraphics[width=0.47\textwidth]{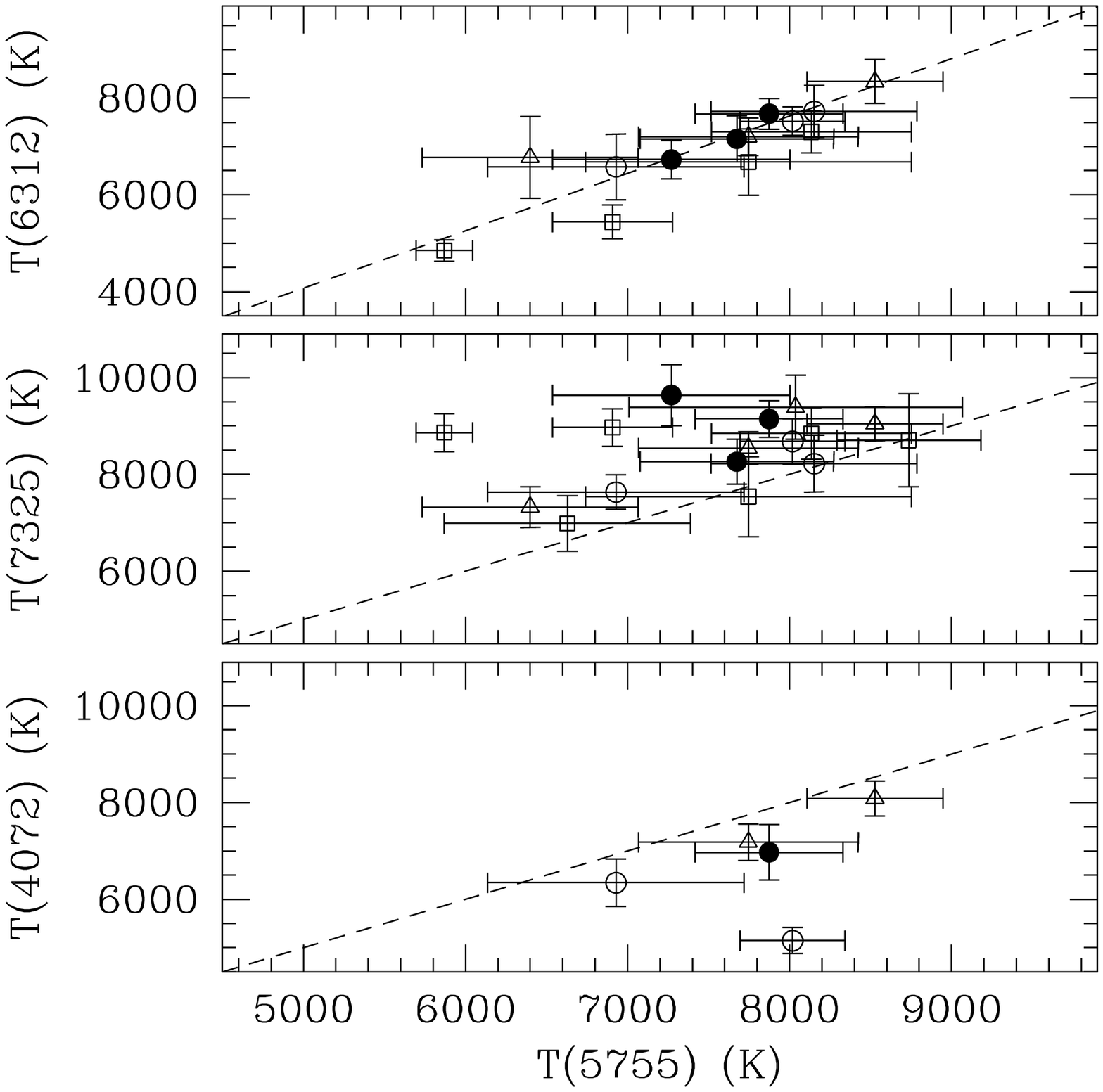} \caption{The
   temperature T(5755) determined from the
   \nii\lin5755/\nii\llin6548,6583 ratio compared with the temperatures
   measured from the auroral lines \siii\lin6312 {\em (top)},
   \oii\lin7325 {\em (middle)} and \sii\lin4072 {\em (bottom)}. In the
   top panel the dashed line represents the relationship predicted by
   the models of \citet{garnett92}, while in the remaining 2 panels the
   line shows the location in the diagrams where T(5755)\,=\,T(7325) and
   T(5755)\,=\,T(4072), respectively. } \label{te} \end{figure}

The empirical relationship found between the various temperatures is
displayed in Fig.~\ref{te}. In the top panel we plot T(5755) against
T(6312). \citet{garnett92} gave simple equations relating electron
temperatures from different ions, based on a 3-zone temperature
stratification of \hii\/ regions. The temperatures T\oii, T\nii\/ and
T\sii\/ are equivalent to the electron temperature in the low-excitation
zone, while T\oiii\/ represents the temperature in the high-excitation
zone. An intermediate-excitation zone is measured by T\siii. The
equations published by \citet{garnett92}, based on photoionization
models by \citet{stasinska82}, are commonly used whenever the data do
not allow the determination of the electron temperature in each
excitation zone:

\begin{center}
\begin{equation}
{\rm T[S\,III] = 0.83\, T[O\,III]\,+\,1700~{\rm K}},
\end{equation}
\end{center}

\begin{center}
\begin{equation}
{\rm T[N\,II] = T[O\,II] = 0.70\, T[O\,III]\,+\,3000~{\rm K}}.
\end{equation}
\end{center}

\noindent Naturally, an empirical verification of these, or equivalent,
equations is highly valuable for extragalactic abundance studies.
Recently \citet{bresolin04}, in their study of metal-rich \hii\/ regions
in the galaxy M51, showed that the predicted T\siii-T\nii\/ relation is
in good agreement with the experimental data. The top panel of
Fig.~\ref{te} shows that good agreement with the model predictions:

\begin{center}
\begin{equation}
{\rm T[S\,III] = 1.19\, T[O\,II]\,-\,1857~{\rm K}},
\end{equation}
\end{center}

\noindent [obtained combining Eq.~(1) and (2)], shown here by the dashed
line, is also found in the current \hii\/ region sample. These results
seem to support the validity of these equations, at least in the
electron temperature range considered (6000--9000~K). On the other hand,
the results of the remaining two comparisons displayed in Fig.~\ref{te}
is less satisfactory. In the 3-zone representation T(5755), T(7325) and
T(4072) should all be representative of the low-excitation zone, and
therefore equivalent. However, T(7325) seems to overestimate the
temperature if compared to T(5755), while the opposite happens for
T(4072). In the case of the \oii\lin7325 line, while it is true that
accounting for a recombination component goes in the right direction to
alleviate the discrepancy (for example using the empirical formula given
by \citealt{liu00}), the effect, when corrected as in
\citet{kennicutt03}, would be negligible in our sample. However, a
correct treatment of recombination should take into account the effect
of temperature gradients within ionized nebulae, and future work on the
importance of the temperature structure of a number of \hii\/ regions in
our sample will shed some light on this important issue. Regarding
T(4072), more observational data need to be collected before confirming
the offset suggested by Fig.~\ref{te}. The reasons for these
discrepancies are thus unclear at the moment, but our results remind us
that, even though the 3-zone representation might be a useful tool for
the interpretation of nebular spectra, it remains a simplification of
the excitation and ionization structure of real \hii\/ regions.

To conclude this section, before we approach the estimate of the
chemical abundances, we must obtain the temperatures required in the
3-zone representation. As a minimum, we need to derive the temperature
of the high-excitation zone, since T\oiii\/ cannot be measured from our
data. This can be done by means of Eqs.~(1) and (2), combining the
results with a weighted mean when both T(6312) and T(5755) are available
[T(7325) was not considered for this estimate]. These two temperatures
also provided \te\/ estimates for the low- and intermediate-excitation
zones, using again Eqs.~(1) and (2) when needed. Finally, for those
\hii\/ regions where only T(7325) was available, we set the
low-excitation temperature equal to T(7325), and derived the high- and
intermediate-excitation zone temperatures from Eqs.~(1) and (2). We have
less confidence in the latter estimates than those obtained from the
availability of both T(6312) and T(5755), because of the results
illustrated in Fig.~\ref{te}. We report in Table~\ref{tetable2} the
adopted electron temperatures thus obtained, and used for determining
the abundances.

%_____________________________________________________________
%                      Table: Electron temperatures measured
%_____________________________________________________________
%
\begin{table*}
\caption{Temperatures measured from auroral lines.} 
\label{tetable}
\centering \renewcommand{\footnoterule}{} 
\begin{tabular}{c c c c c}     % 5 columns 
\hline\hline
   & T(7325) & T(5755) & T(6312) & T(4072) \\ 
\phantom{aaa}ID\phantom{aaa}    & (K) & (K) & (K) & (K)\\[0.5mm]
\hline\\
\input{te}
\hline
\end{tabular}
\end{table*}
%_____________________________________________________________
%

%_____________________________________________________________
%                                                             Figure: Te
%-------------------------------------------------------------

%_____________________________________________________________
%

%_____________________________________________________________
%                      Table: Electron temperatures adopted
%_____________________________________________________________
%
\begin{table}
\caption{Adopted temperatures for the 3-zone representation.} 
\label{tetable2}
\centering \renewcommand{\footnoterule}{} 
\begin{tabular}{c c c c }     % 4 columns 
\hline\hline
   & T(O$^+$, N$^+$, S$^+$) & T(S$^{+2}$) & T(O$^{+2}$) \\ 
\phantom{aaa}ID\phantom{aaa}    & (K) & (K)  & (K)\\[0.5mm]
\hline\\
\input{te_adopted}
\hline
\end{tabular}
\end{table}
%_____________________________________________________________
%

%__________________________________________________________________

\section{Chemical abundance: direct method} The temperatures derived in
the previous section can now be used to measure the nebular chemical
abundances, keeping in mind, however, the caution expressed at the
beginning of Sect.~4 regarding the abundance biases in metal-rich \hii\/
regions. With the 3-zone representation we have derived ionic
abundances, adopting the electron temperature of the low-excitation zone
for O$^+$, N$^+$ and S$^+$, the temperature of the
intermediate-excitation zone for S$^{+2}$, and the temperature of the
high-excitation zone for O$^{+2}$ (see Table~\ref{tetable2}). In order
to compute total element abundances, we then made the common
assumptions: O/H = (O$^+$\,+\,O$^{+2}$)/H$^+$, N/O = N$^+$/O$^+$, while
for S/O we have used the ionization correction formula of
\citet{stasinska78}, as used in \citet{bresolin04} and
\citet{kennicutt03}. The abundances of oxygen relative to hydrogen and
of nitrogen and sulphur relative to oxygen thus derived are reported in
Table~\ref{abundances}.

The reader should bear in mind that these abundances will be checked
against a more detailed analysis, to be presented in a forthcoming
paper, to which we postpone the report on the detailed abundance
properties of our \hii\/ region sample. In this section we briefly
summarize the trends of the S/O and N/O abundance ratios with O/H, in
order to  characterize our sample and make a comparison with works in
the literature. The variation of heavy element ratios, in particular
N/O, with metallicity offers a crucial insight into the nucleosynthetic
nature of these elements (\citealt{henry00}), and it is therefore
important to extend the measurements to metal-rich environments, such as
those encountered in the central regions of spiral galaxies
(\citealt{bresolin04}, \citealt{garnett04b}).

\begin{figure}
   \centering

   \includegraphics[width=0.47\textwidth]{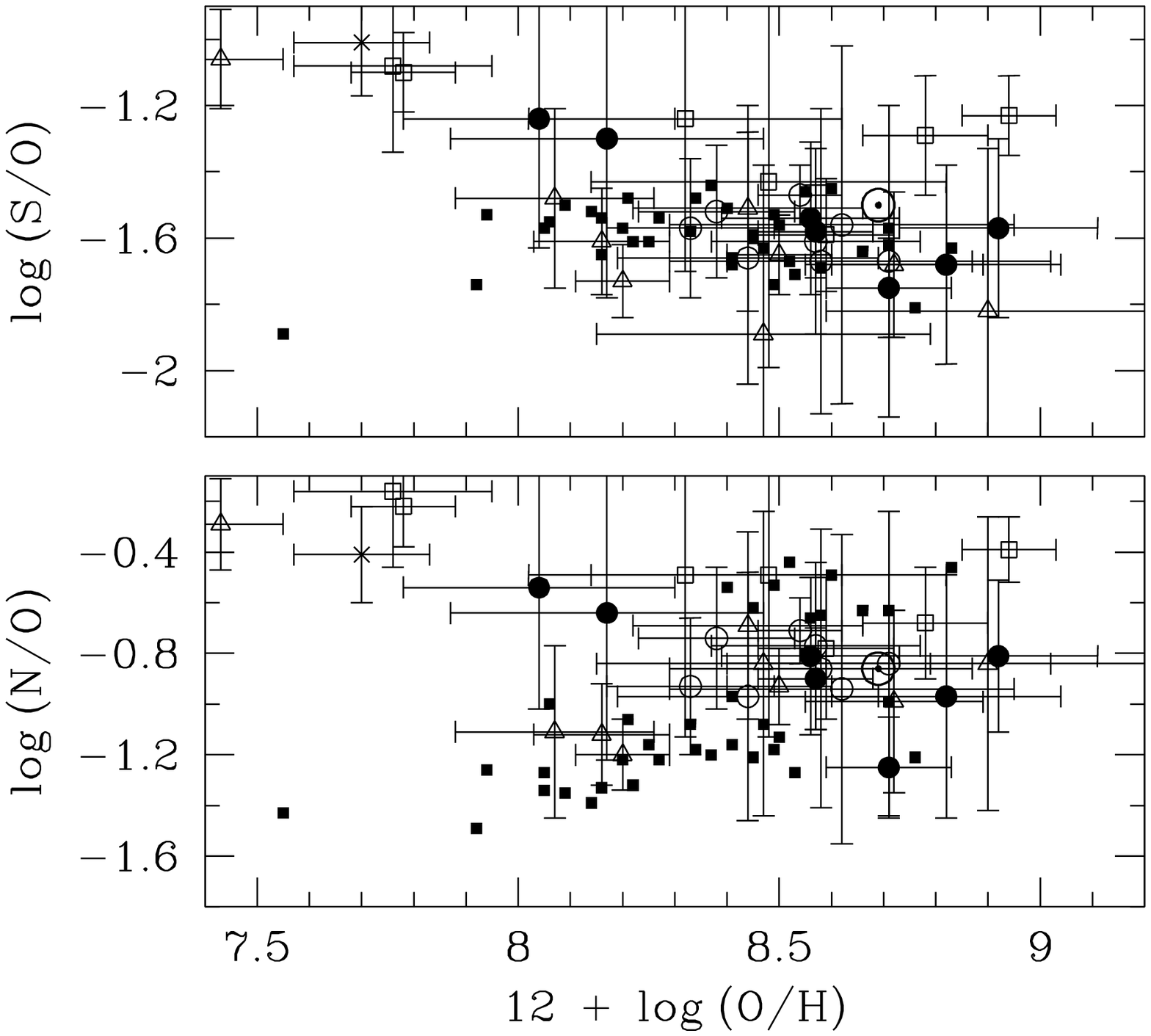} \caption{The S/O
   {\em (top)} and N/O {\em (bottom)} abundance ratio trends with O/H
   for all objects in Table~\ref{abundances}. A comparison sample, drawn
   from \citet[NGC~2403]{garnett97}, \citet[M101]{kennicutt03} and
   \citet[M51]{bresolin04}, is shown by small full square symbols. The
   solar values, indicated by the $\odot$ symbol, are taken from
   \citet{lodders03}. } \label{sulfur} \end{figure}

\begin{figure}
   \centering

   \includegraphics[width=0.47\textwidth]{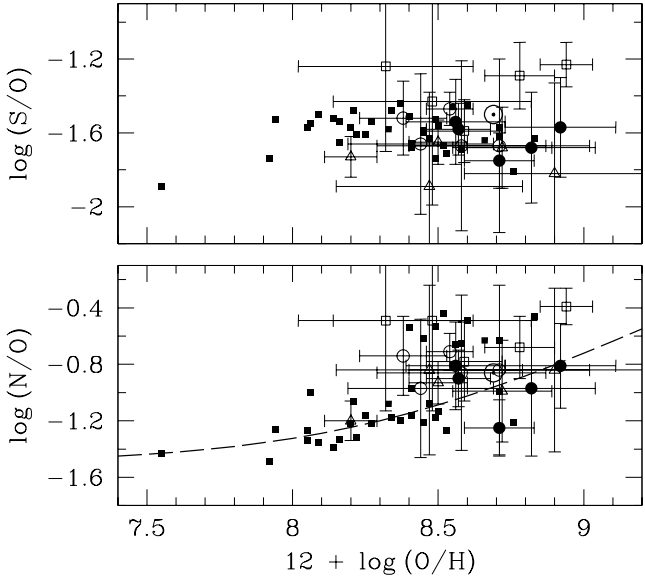} \caption{Same as
   Fig.~\ref{sulfur}, but including only nebulae from the VLT sample
   where the electron temperature has been computed from the
   availability of at least one of the \nii\lin5755 and \siii\lin6312
   auroral lines (objects marked by asterisks in
   Table~\ref{abundances}). The dashed line represents a simple model,
   in which a primary nitrogen component is superposed on a secondary
   component, which is proportional to O/H. } \label{sulfur2}
   \end{figure}

The  S/O and N/O ratios of all objects included in
Table~\ref{abundances} are plotted as a function of O/H in
Fig.~\ref{sulfur}, where we add a comparison sample of extragalactic
\hii\/ regions with published \te-based abundances, extracted from
\citet[NGC~2403]{garnett97}, \citet[M101]{kennicutt03} and
\citet[M51]{bresolin04}, and shown by the small full square symbols. The
objects from our new observations, indicated by the usual symbols
(defined in Fig.~\ref{compare_lines}) and the corresponding error bars,
are generally consistent with the known trends of roughly constant S/O
[$\log$(S/O)\,$\simeq$\,$-1.6$] and N/O increasing with O/H in the
high-metallicity regime, although a number of outliers are clearly
present at the low-abundance end. This is likely due to the inadequacy
of the inferred temperatures for the 3-zone representation in those
cases where, among the auroral lines, only \oii\lin7325 was measured. In
fact, the abundances for the comparison sample of \hii\/ regions in
NGC~2403, M101 and M51 were derived from the measurement of \nii\lin5755
and \siii\lin6312 in their spectra, while disregarding abundances based
on the \oii\lin7325 auroral line. As shown in Sect.~4, T(7325) appears
to overestimate the electron temperature in the low-excitation zone,
thus leading to an underestimate of the oxygen abundance. If we limit
the diagram to include only those objects in the VLT sample where
\nii\lin5755 and/or \siii\lin6312 are available (marked by asterisks in
Table~\ref{abundances}), which arguably allows a more robust application
of the 3-zone model, a picture which is more consistent with the
previous abundance works emerges, as seen in Fig.~\ref{sulfur2}. In the
bottom panel of this figure we have also drawn as a reference (dashed
line) the simple model for N/O introduced by \citet{kennicutt03} as the
sum of a primary, constant component [$\log$(N/O)\,$=$\,$-1.5$] and a
secondary component, for which N/O is proportional to O/H
[$\log$(N/O)\,$=$\,$\log$(O/H)\,+\,2.2], which reproduces fairly well
the metallicity dependence of N/O in the \hii\/ regions of M101. The
scatter in N/O at constant oxygen abundance is well-known (see
\citealt{henry00}), so it is not surprising to find objects deviating
(at the 1-2\,$\sigma$ level) from the dashed line.

Among the objects included in Table~\ref{abundances}, we draw the
attention to a few interesting cases. First of all, NGC~1232-11, which
is characterized by peculiar emission line ratios (e.g. large
\oi\lin6300/H$\beta$) and which appears to deviate from the \hii\/
region sequence in Fig.~\ref{baldwin}, has also a much higher \te\/ than
the rest of the sample, and a correspondingly small O/H for its inner
position in the galaxy. The wavelengths of its emission lines are not
discordant from those of the remaining \hii\/ regions in NGC~1232,
therefore it is not a background emission-line galaxy at larger
redshift.

The oxygen abundance derived for NGC~1232-07,
12\,+\,log(O/H)\,=\,8.9\,$\pm$\,0.3, is in good agreement with the value
of 8.95\,$\pm$\,0.20 reported by \citet[their object
CDT1]{castellanos02}. At the time of their publication, this object was
the most metal-rich extragalactic \hii\/ region with an electron
temperature measured from auroral lines. In our VLT sample, the most
metal-rich nebulae do not exceed the oxygen abundance of this \hii\/
region. In particular, for NGC~5236-11, in the very nucleus of the M83
galaxy, we find an abundance 12\,+\,log(O/H)\,=\,8.94\,$\pm$\,0.09,
while for NGC~2997-13 we find  12\,+\,log(O/H)\,=\,8.92\,$\pm$\,0.19.
Therefore, with the direct method adopted in this work, applied to
observations obtained at the VLT, we have not been able to find
abundances larger than about 1.6 times the solar one
[12\,+\,log(O/H)$_\odot$\,=\,8.69]. This conclusion, however, is likely
to be revised (in either direction) if biases due to temperature
stratification (\citealt{stasinska05}) are duely taken into account.

\begin{figure}
   \centering

   \includegraphics[width=0.47\textwidth]{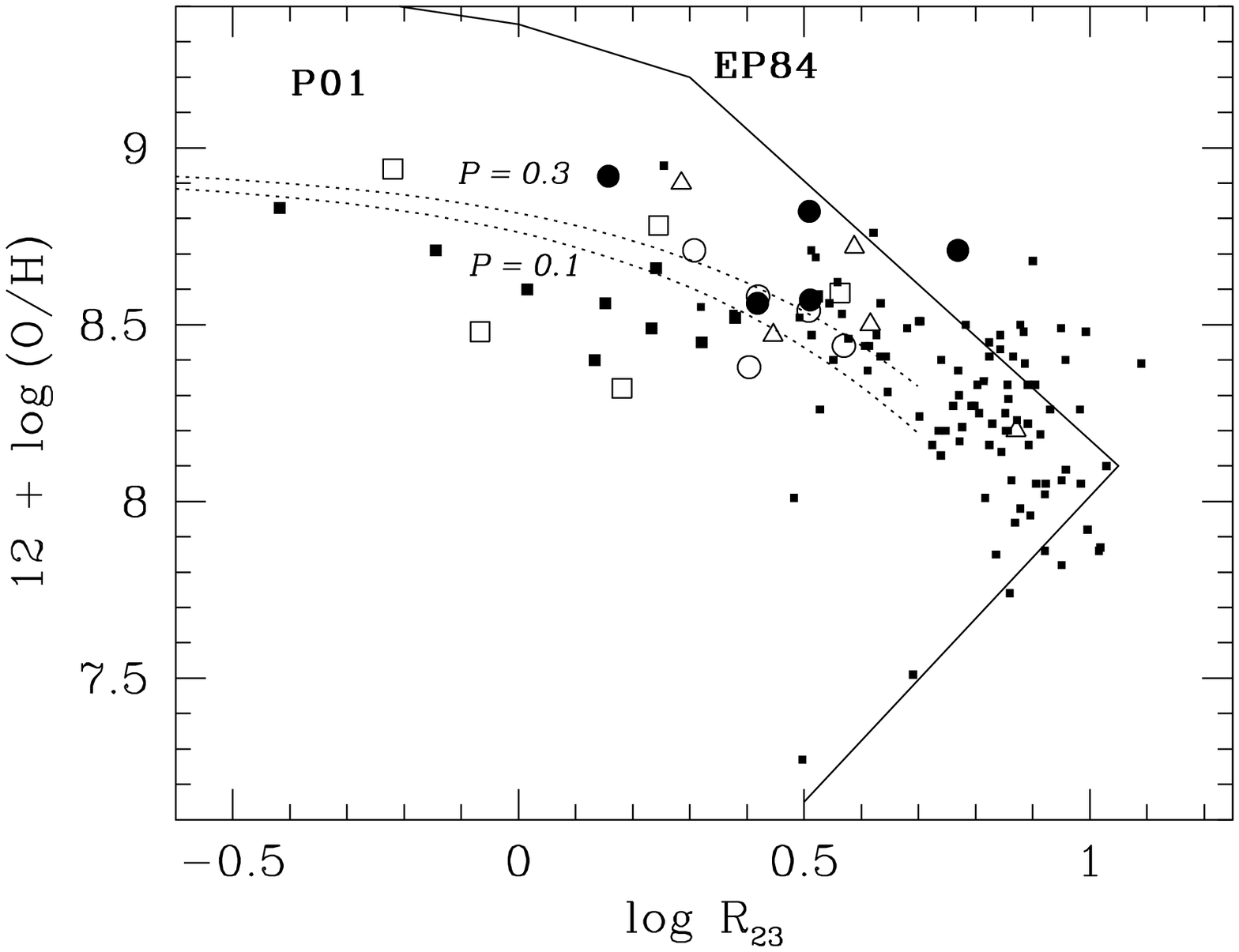} \caption{The
   abundance indicator $R_{23}$ as a function of oxygen abundance,
   including all \hii\/ regions with measured \nii\lin5755 or
   \siii\lin6312 auroral lines. The comparison sample (small squares) is
   the same as in Fig.~\ref{sulfur}. The continuous and dotted lines
   show, respectively, the $R_{23}$ calibrations by \citet{edmunds84}
   (EP84) and \citet{pilyugin01} (P01). The latter has been drawn for
   two different values of the excitation parameter, $P=0.1$ and
   $P=0.3$. This range encompasses the majority of the \hii\/ regions in
   the VLT sample. } \label{r23} \end{figure}

\begin{figure}
   \centering

   \includegraphics[width=0.47\textwidth]{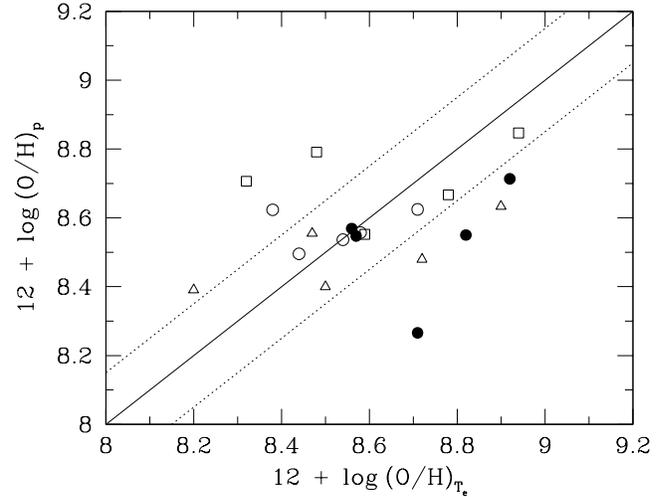}
   \caption{Comparison between direct (\te-based) abundances  and those
   estimated via the $R_{23}$ calibration of \citet{pilyugin01}. The
   dotted lines are drawn 0.15 dex below and above the line of equal value.}
      \label{direct_pily} \end{figure}

To conclude this preliminary look at the abundance properties of our
sample, we plot in Fig.~\ref{r23} the indicator $R_{23}$ as a function
of the \te-based oxygen abundance, again including only objects with
\nii\lin 5755 and/or \siii\lin6312 detections. In this diagram we also
show the points corresponding to the \hii\/ regions in NGC~2403, M101
and M51 from the papers mentioned above (small full square symbols). The
two widely used $R{23}$ calibrations of \citet{edmunds84} and
\citet{pilyugin01} (the latter applicable for 12\,+\,log(O/H)\,$>$\,8.2,
according to the latter author) are shown by the continuous and dotted
lines, respectively. The \citet{pilyugin01} calibration attempts to
account for the sensitivity of $R_{23}$ to the ionization parameter, by
introducing the quantity
$P$\,=\,\oiii\llin4959,5007/(\oii\lin3727\,+\,\oiii\llin4959,5007). Two
curves, corresponding to $P=0.1$ and $P=0.3$, the same values used in
Fig.~\ref{statistical} to bracket most of the \hii\/ regions in the
current sample, are drawn in Fig.~\ref{r23}. As can be seen, the most
metal-rich \hii\/ regions, in particular those in NGC~5236 (open
squares), reach values of $R_{23}$ that are comparable to those found in
M51 \hii\/ regions by \citet{bresolin04}, and have similar O/H
abundances. Fig.~\ref{r23} confirms earlier findings
(\citealt{pindao02}, \citealt{kennicutt03}, \citealt{bresolin04}) that
indicated how some of the calibrations of statistical methods available
in the literature (e.g.~\citealt{edmunds84}, \citealt{zaritsky94}) can
severely overestimate the abundance of metal-rich \hii\/ regions, while
others (e.g.~\citealt{pilyugin01}) might be less affected by systematic
differences compared to direct abundances, even though the two methods
can still give significantly discrepant results for individual \hii\/
regions. This is shown in Fig.~\ref{direct_pily}, where we compare the
\te-based abundances with those estimated from the \citet{pilyugin01}
$R_{23}$ calibration. The dotted lines are drawn 0.15 dex above and
below the line of equal value (full line), to aid in the comparison with
a similar diagram presented by \citet[their Fig.~15]{pilyugin04}. For
our metal-rich sample we clearly find a larger scatter than found by
these authors.

%_____________________________________________________________
%                      Table: Abundances
%_____________________________________________________________
%
\begin{table}
\begin{minipage}[t]{0.4\textwidth}
\caption{Abundance estimates from the 3-zone representation.} 
\label{abundances}
\centering \renewcommand{\footnoterule}{} 
\begin{tabular}{c c c c }     % 4 columns 
\hline\hline
ID \phantom{*}  &  12 + log (O/H) & log (N/O)  & log (S/O)\\[0.5mm]
\hline\\
\input{abund.tex}
\hline
\end{tabular}
\begin{flushleft}
{\footnotesize NOTES: for the objects marked by an asterisk the abundances have 
been derived from the measurement of either \nii\lin5755, \siii\lin6312, or both.}
\end{flushleft}
\end{minipage}
\end{table}
%_____________________________________________________________
%

%_____________________________________________________________
%                                                     Figure: S/O, N/O
%-------------------------------------------------------------

%_____________________________________________________________
%

%_____________________________________________________________
%                                               Figure: S/O, N/O (2)
%-------------------------------------------------------------

%_____________________________________________________________
%

%_____________________________________________________________
%                                               Figure: R23
%-------------------------------------------------------------

%_____________________________________________________________
%

%_____________________________________________________________
%                                  Figure: directs vs Pilyugin
%-------------------------------------------------------------

%_____________________________________________________________
%

\section{Conclusions} We have presented new optical VLT spectroscopy for
a sample of extragalactic \hii\/ regions, comprising about 70 nebulae,
distributed in five galaxies: NGC~1232, NGC~1365, NGC~2903, NGC~2997 and
NGC~5236. The target galaxies were selected in order to maximize the
odds of obtaining spectra of truly metal-rich \hii\/ regions, with
oxygen abundances around the solar value and above. Our principal goal
in this first paper of a series was to present the emission line
measurements, in particular for the 32 objects where we have been able
to detect auroral lines for different ions. With the aid of these lines,
and adopting a 3-zone description for the excitation
structure of the nebulae, we have derived electron temperatures and
abundances of O, N and S, neglecting the abundance biases that are
likely to be introduced by the temperature stratification of the
nebulae. The impact of these biases on the chemical abundance
measurements presented here will be assessed in a future publication.

The direct (\te-based) method of abundance determination has provided
only a handful of objects of genuine high metallicity, that is well
above solar, up to 12\,+\,log(O/H)\,$\simeq$\,8.9. We have measured a
direct abundance for two additional \hii\/ regions, besides the CDT1
nebula studied by \citet{castellanos02}, where the oxygen abundance
reaches this value: our NGC~2997-13 and NGC~5236-11. Of course, this
result does not exclude the presence of \hii\/ regions of higher
metallicity in these galaxies, but it is interesting to note that one of
these objects, NGC~5236-11, lies at the center, i.e. where we expect the
oxygen abundance to be highest, of M83, a galaxy which has been known to
be among the most metal-rich spirals for a long time. The \te-based
oxygen abundance of an \hii\/ region near  the center of M51, another
metal-rich spiral galaxy, was found by \citet{bresolin04} to exceed by
only 40\%  the solar oxygen abundance. It thus appears conceivable
that we have started to measure electron temperatures among the most
metal-rich \hii\/ regions in spiral galaxies. Deep spectroscopy of a
larger number of \hii\/ regions within the same galaxies studied here
might provide better constraints on the metallicity at the top-end of
the scale.

What appears to be well established is that at high metallicity the
direct abundances are systematically smaller than the abundances derived
from most statistical methods calibrated by means of photoionization
models. We confirm earlier results that provided some of the first solid
empirical evidence for this discrepancy (\citealt{kennicutt03},
\citealt{garnett04}, \citealt{bresolin04}). With the availability of the
new direct measurements provided in these works, some of the existing
calibrations for statistical methods appear inconsistent with the direct
measurements at high metallicity. A thorough analysis of abundance
calibrators taking into account strong electron temperature
stratification at high metallicities and additional observational data
(e.g.~from infrared fine structure lines) is however needed. The
widespread  use of strong line indicators in estimating the chemical
abundances of star-forming regions both at low and high redshift makes
this an obviously important issue.

%%%%%%%%%%%%%%%%%%%%%%%%%%%%%%%%%%%%%%%%%%%%%%%%
%%%%%%%%%%%%%%%%%%%%%%%%%%%%%%%%%%%%%%%%%%%%%%%%
%\begin{acknowledgements}
%
%\end{acknowledgements}

\appendix
\section{Serendipitous discovery of a $z\simeq2.55$ QSO}

\begin{figure}
\centering

\includegraphics[width=0.47\textwidth]{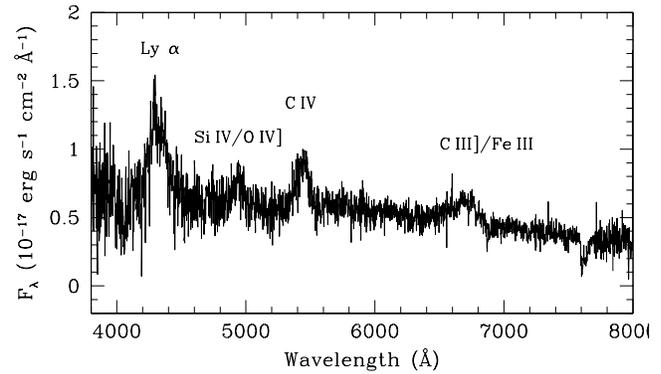}
\caption{The spectrum of a QSO from our MOS setup in NGC~1365.
}
\label{qso}
\end{figure}

The spectrum of our target object for slitlet 2 in the NGC~1365 MOS
setup turned up to be that of a QSO, instead of a star-forming region
within this galaxy. The position relative to the galaxy center is
($-$26$\arcsec$, 182$\arcsec$), corresponding to RA\,=\,03$^\mathrm{h}$
33$^\mathrm{m}$ 34\fs2, DEC\,=\,$-$36$\degr$ 05$\arcmin$ 23$\farcs$8.
This object is marked by the open circle at the top of
Fig.~\ref{ngc1365}. By convolving the flux-calibrated spectrum with the
response function of broad-band filters in the Johnson photometric
system, we have derived $V=21.9$ and $B-V=0.4$. The broad lines detected
in the spectrum (see Fig.~\ref{qso}) have been used to derive a redshift
$z\simeq2.55$.

%______________________________________________________________

%\clearpage

\end{document}

%% file: ngc1232_global.tex
02 &     $-38$ &   $176$ &   199 & 2.2  & 0.16 $\pm$ 0.10 & 140 & 7.8 & 0.0  \\
03 &     $-69$ &   $150$ &   178 & 2.2  &  0.00 $\pm$ 0.05 & 76  & 4.7 & 0.0  \\
04 &     $135$ &   $118$ &   199 & 2.1  &  0.44 $\pm$ 0.10 & 296 & 40.1 & 2.0  \\
05 &     $ 20$ &   $ 93$ &   109 & 3.9  &  0.39 $\pm$ 0.05 & 109 & 90.7 & 1.0  \\
06 &     $ 65$ &   $ 80$ &   116 & 2.7  &  0.28 $\pm$ 0.05 & 57 & 25.9 & 0.6  \\
07 &     $ 59$ &   $ 60$ &    94 & 2.0  &  0.25 $\pm$ 0.05 & 57 & 78.2 & 0.3  \\
08 &     $ 16$ &   $ 28$ &    37 & 3.3  &  0.46 $\pm$ 0.08 & 62 & 12.9 & 0.0  \\
09 &     $ 60$ &   $  7$ &    62 & 4.9  &  0.40 $\pm$ 0.10 & 29 & 23.5 & 0.0  \\
10 &     $ 75$ &   $-10$ &    76 & 3.1  &  0.44 $\pm$ 0.10 & 83 & 28.0 & 0.2  \\
\phantom{$^d$}11\footnote{second object in slitlet 10} &     $ 75$ &   $-14$ &    77 & 1.8  &  0.44 $\pm$ 0.05 & 176 & 6.2 & 0.0  \\
12 &     $ 47$ &   $-57$ &    77 & 2.7  &  0.50 $\pm$ 0.10 & 62 & 9.1 & 0.0  \\
14 &     $  4$ &   $-99$ &   111 & 4.2  &  0.02 $\pm$ 0.05 & 174 & 85.2 & 2.0  \\
15 &     $102$ &   $-129$ &   172 & 3.4  &  0.07 $\pm$ 0.05 & 180 & 6.9 & 2.0  \\

%% file: ngc1365_global.tex
03 &     $-$14 &   162 &   221 & 2.0  &   0.35 $\pm$  0.10 & 204 & 14.6 & 2.5   \\
04 &     $-$43 &   122 &   202 & 1.7  &   0.26 $\pm$  0.05 & 15 & 5.3 & 1.8   \\
05 &     $-$49 &   105 &   189 & 4.2  &   0.40 $\pm$  0.05 & 48 & 92.3  & 0.3  \\
06 &     $-$57 &    77 &   168 & 2.3  &   0.36 $\pm$  0.05 & 28 & 16.0 & 0.5   \\
07 &    $-$111 &    52 &   224 & 2.8  &   0.05 $\pm$  0.05 & 68 & 7.5 & 2.1   \\
08 &      66 &    55 &    95   & 3.8  &   0.38 $\pm$  0.05 & 59 & 105.4 & 0.3   \\
09 &      17 &    17 &    25   & 4.3  &   0.25 $\pm$  0.10 & 42 & 177.9 & 4.8   \\
10 &      79 &    18 &   119   & 1.9  &   0.55 $\pm$  0.10 & 44 & 31.2 & 0.0   \\
11 &      87 &    $-$6 &   147 & 1.8  &   1.06 $\pm$  0.10 & 13 & 6.7 & 0.1   \\
12 &      86 &   $-$24 &   159 & 1.1  &   0.69 $\pm$  0.10 & 43 & 25.8 & 0.0   \\
13 &      76 &   $-$63 &   180 & 2.7  &   0.48 $\pm$  0.15 & 56 & 24.4 & 0.0   \\
14 &      79 &   $-$75 &   196 & 3.9  &   0.61 $\pm$  0.05 & 50 & 44.0 & 0.4   \\
15 &      48 &  $-$105 &   188 & 3.7  &   0.43 $\pm$  0.10 & 195 & 109.6  & 0.0  \\
16 &      27 &  $-$141 &   207 & 1.9  &   0.20 $\pm$  0.05 & 40 & 20.8  & 0.0  \\
17 &      10 &  $-$164 &   221 & 1.9  &   0.03 $\pm$  0.03 & 98 & 11.6 & 1.1   \\

%% file: ngc2903_global.tex
02  &     14 &   178 &   192 & 4.6    &  0.05  $\pm$  0.05 & 31  & 14.7 & 0.5  \\
03  &    $-$12 &   159 &   190 & 3.4    &   0.03  $\pm$ 0.05 & 100 & 16.1 & 2.9  \\
04  &     55 &   123 &   138 &  3.1   &   0.60  $\pm$  0.10 & 27  & 13.7 & 0.2  \\
05  &     75 &    78 &   139 &  2.9    &  0.31  $\pm$  0.10 & 77  & 51.2 & 1.2  \\
06  &    $-$22 &   100 &   137 & 2.0     &  0.28  $\pm$  0.06 & 217 & 28.5 & 1.2  \\
07  &     $-$5 &    64 &    77 &  4.9    &  1.04  $\pm$  0.10 & 28  & 128.2 &  0.8 \\
08  &     23 &    38 &    48 & 2.9    &   0.39  $\pm$  0.10 & 43  & 118.9 & 2.5  \\
12  &     43 &   $-$78 &   145 & 2.4     &  0.39  $\pm$  0.05 & 43  & 66.9 & 0.2  \\
14  &    $-$64 &   $-$95 &   129 & 2.0     &  0.49  $\pm$  0.15 & 1   & 10.6 & 0.0  \\

%% file: ngc2997_global.tex
03 &      43 &   154 &   212 & 2.7   &  0.10 $\pm$  0.10 & 97 & 15.3 & 0.0  \\
04 &      $-$6 &   153 &   196 & 4.4   &   0.27 $\pm$  0.10 & 121 & 26.7 & 0.0  \\
05 &     $-$36 &   128 &   163 & 3.9    &  0.14 $\pm$  0.10 & 147 & 54.9 & 3.7  \\
06 &     $-$52 &   113 &   146 & 4.6   &   0.28 $\pm$  0.05 & 151 & 115.3 & 3.7  \\
07 &      81 &    57 &   122 & 3.2    &  0.99 $\pm$  0.10 & 77 & 59.6 & 0.0  \\
08 &     106 &    22 &   119 & 2.2    &  1.00 $\pm$  0.10 & 23 & 13.5 & 0.4  \\
09 &     131 &    $-$8 &   135 & 5.3   &   0.27 $\pm$  0.05 & 34 & 27.9 & 0.2  \\
10 &    $-$112 &    28 &   116 & 4.2    &  0.07 $\pm$  0.05 & 51 & 38.4 & 2.1  \\
11 &    $-$128 &    16 &   131 & 3.8   &   0.18 $\pm$  0.04 & 95 & 13.9 & 1.8  \\
12 &       9 &   $-$50 &    64 & 1.9   &   0.76 $\pm$  0.07 & 42 & 13.1 & 1.8  \\
13 &     $-$14 &   $-$69 &    93 & 2.8   &   0.56 $\pm$  0.05 & 96 & 49.5 & 2.2  \\
14 &    $-$156 &   $-$54 &   188 & 4.3   &   0.40 $\pm$  0.05 & 189 & 27.4 & 4.5  \\
15 &     $-$43 &  $-$110 &   157 & 3.0   &   0.67 $\pm$  0.05 & 51 & 22.2 & 2.0  \\
16 &     $-$13 &  $-$126 &   166 & 3.9   &   0.41 $\pm$  0.00 & 52 & 14.0 & 1.5  \\

%% file: ngc5236_global.tex
02  &    $-$13 &   193 &   207 & 1.8   &   0.58 $\pm$  0.05 & 23  & 26.5 & 0.6  \\
03  &    117 &   165 &   203 & 1.5   &   0.53 $\pm$  0.10 & 102 & 98.6 & 0.0  \\
04  &     50 &   151 &   163 & 2.8   &   0.35 $\pm$  0.05 & 26 & 123.1 & 1.0  \\
05  &    136 &   124 &   184 & 2.6   &   0.70 $\pm$  0.10 & 45 &  130.0 & 0.0  \\
06  &   $-$126 &   108 &   186 & 1.5   &   0.14 $\pm$  0.05 & 47 &  76.2 & 2.0  \\
07  &   $-$128 &    76 &   166 & 1.7    &  0.13 $\pm$  0.10 & 13 &  71.0 & 1.5  \\
08  &     48 &    58 &    75 & 2.3   &   0.68 $\pm$  0.08 & 12 &  142.0 & 0.9  \\
09  &      7 &    30 &    32 & 1.7   &   1.12 $\pm$  0.10 & 46 &  137.8 & 0.6  \\
10  &   $-$107 &     8 &   115 & 2.6    &  0.29 $\pm$  0.10 & 27 &  175.4 & 1.2  \\
11  &     $-$5 &    $-$1 &     5 & 1.6   &   0.27 $\pm$  0.05 & 48 &  1264.6 & 2.6  \\
12  &    100 &   $-$33 &   116 & 3.1   &   0.48 $\pm$  0.10 & 64 &  167.6 & 0.5  \\
13  &   $-$104 &   $-$54 &   119 & 2.5   &   0.34 $\pm$  0.10 & 55 &  212.6 & 1.3  \\
14  &    $-$78 &   $-$73 &   107 & 1.9    &  0.15 $\pm$  0.08 & 31 &  166.1 & 1.5  \\
15  &     65 &  $-$109 &   141 & 1.4   &   0.22 $\pm$  0.08 & 7  &  47.5 & 1.9  \\
16  &     33 &  $-$125 &   141 & 2.2   &   0.61 $\pm$  0.05 & 74 &  148.6 & 1.3  \\
17  &    $-$22 &  $-$151 &   159 & 2.3   &   0.36 $\pm$  0.15 & 35 &  105.1 & 0.3  \\
18  &    $-$48 &  $-$173 &   185 & 2.6   &   0.57 $\pm$  0.10 & 39 &  29.8 & 0.6  \\
19  &     70 &  $-$188 &   221 & 2.5    &  0.60 $\pm$  0.05  & 32 &  11.8 & 0.0  \\

%% file: ngc1232_b.tex
  02 &  391 $\pm$   33 &     \nodata   &   26 $\pm$    2 &   \nodata       &   41 $\pm$    3 &      \nodata  &   44 $\pm$    3 &  130 $\pm$    8 &   \nodata       &   \nodata       & 11.9 $\pm$  1.2 \\
  03 &  333 $\pm$   25 &25.3 $\pm$ 3.5 &   29 $\pm$    4 &   \nodata       &   47 $\pm$    4 &      \nodata  &  100 $\pm$    8 &  298 $\pm$   21 & 8.60 $\pm$ 3.27 &   \nodata       & 14.0 $\pm$  1.6 \\
  04 &  207 $\pm$   17 &29.1 $\pm$ 2.4 &   28 $\pm$    2 & 1.95 $\pm$ 0.36 &   45 $\pm$    2 &  4.3 $\pm$  0.4 &  135 $\pm$    8 &  401 $\pm$   24 &   \nodata       &   \nodata       & 10.6 $\pm$  0.9 \\
  05 &  253 $\pm$   16 & 4.8 $\pm$ 0.6 &   25 $\pm$    1 & 2.88 $\pm$ 0.29 &   46 $\pm$    2 &  3.9 $\pm$  0.3 &   34 $\pm$    2 &  100 $\pm$    6 & 2.06 $\pm$ 0.23 & 0.54 $\pm$ 0.15 & 11.9 $\pm$  0.8 \\
  06 &  211 $\pm$   14 &     \nodata   &   25 $\pm$    1 &   \nodata       &   46 $\pm$    2 &  3.0 $\pm$  0.8 &   18 $\pm$    1 &   50 $\pm$    3 & 2.19 $\pm$ 0.82 & 0.66 $\pm$ 0.26 &  9.3 $\pm$  0.7 \\
  07 &  163 $\pm$   10 &     \nodata   &   23 $\pm$    1 &   \nodata       &   46 $\pm$    2 &  3.1 $\pm$  0.3 &    7 $\pm$    1 &   22 $\pm$    1 & 1.91 $\pm$ 0.23 & 0.39 $\pm$ 0.16 &  9.7 $\pm$  0.6 \\
  08 &   58 $\pm$    5 &     \nodata   &   24 $\pm$    2 &   \nodata       &   47 $\pm$    3 &      \nodata  &    3 $\pm$    1 &    6 $\pm$    1 &   \nodata       &   \nodata       &  7.0 $\pm$  1.0 \\
  09 &   87 $\pm$    7 &     \nodata   &   23 $\pm$    2 &   \nodata       &   47 $\pm$    3 &      \nodata  &    3 $\pm$    1 &    8 $\pm$    1 &   \nodata       &   \nodata       & 10.2 $\pm$  1.2 \\
  10 &  134 $\pm$   11 & 2.4 $\pm$ 1.3 &   25 $\pm$    1 &   \nodata       &   49 $\pm$    3 &  4.1 $\pm$  0.5 &    8 $\pm$    1 &   23 $\pm$    1 & 2.35 $\pm$ 0.39 &   \nodata       &  9.3 $\pm$  0.8 \\
  11 &  337 $\pm$   23 &15.4 $\pm$ 2.6 &   25 $\pm$    3 &25.84 $\pm$ 3.20 &   47 $\pm$    3 &  8.0 $\pm$  1.1 &   20 $\pm$    2 &   63 $\pm$    4 &12.01 $\pm$ 1.21 &   \nodata       & 11.7 $\pm$  1.2 \\
  12 &  276 $\pm$   24 &     \nodata   &   26 $\pm$    3 &   \nodata       &   45 $\pm$    4 &      \nodata  &   17 $\pm$    3 &   39 $\pm$    4 &   \nodata       &   \nodata       & 12.2 $\pm$  2.4 \\
  14 &  305 $\pm$   19 & 4.1 $\pm$ 0.4 &   26 $\pm$    1 & 5.14 $\pm$ 0.38 &   46 $\pm$    2 &  3.2 $\pm$  0.4 &   27 $\pm$    2 &   80 $\pm$    5 & 2.09 $\pm$ 0.20 & 0.84 $\pm$ 0.12 & 11.4 $\pm$  0.7 \\
  15 &  458 $\pm$   32 &     \nodata   &   25 $\pm$    2 &   \nodata       &   46 $\pm$    3 &      \nodata  &   35 $\pm$    3 &  103 $\pm$    7 & 6.19 $\pm$ 2.32 &   \nodata       &  9.1 $\pm$  1.0 \\

%% file: ngc1232_r.tex
  02 &      \nodata  &   \nodata       &   17 $\pm$    2 &  286 $\pm$   28 &   47 $\pm$    5 &  3.2 $\pm$  0.7 &   26 $\pm$    3 &   19 $\pm$    2 &  6.1 $\pm$  0.9 &  8.4 $\pm$  1.6 &   26 $\pm$    5 \\
  03 &  7.0 $\pm$  1.3 &   \nodata       &   16 $\pm$    2 &  288 $\pm$   23 &   43 $\pm$    4 &      \nodata  &   34 $\pm$    3 &   23 $\pm$    2 &  7.2 $\pm$  1.3 &      \nodata  &   20 $\pm$    4 \\
  04 &  2.0 $\pm$  0.2 & 1.47 $\pm$ 0.19 &    7 $\pm$    1 &  276 $\pm$   26 &   22 $\pm$    2 &  3.3 $\pm$  0.4 &   14 $\pm$    1 &    9 $\pm$    1 &  8.2 $\pm$  0.9 &  4.0 $\pm$  0.5 &   19 $\pm$    4 \\
  05 &  3.0 $\pm$  0.2 & 0.66 $\pm$ 0.11 &   28 $\pm$    2 &  287 $\pm$   20 &   83 $\pm$    6 &  3.6 $\pm$  0.3 &   34 $\pm$    2 &   25 $\pm$    2 &  6.5 $\pm$  0.5 &  3.4 $\pm$  0.3 &   25 $\pm$    5 \\
  06 &  3.7 $\pm$  0.4 &   \nodata       &   30 $\pm$    2 &  287 $\pm$   20 &   90 $\pm$    6 &  2.9 $\pm$  0.4 &   44 $\pm$    3 &   31 $\pm$    2 &  3.7 $\pm$  0.4 &  3.2 $\pm$  0.4 &    3 $\pm$    1 \\
  07 &  2.1 $\pm$  0.2 & 0.36 $\pm$ 0.15 &   36 $\pm$    3 &  286 $\pm$   20 &  120 $\pm$    8 &  2.7 $\pm$  0.2 &   38 $\pm$    3 &   24 $\pm$    2 &  2.6 $\pm$  0.2 &  1.4 $\pm$  0.2 &   17 $\pm$    3 \\
  08 &      \nodata  &   \nodata       &   22 $\pm$    2 &  292 $\pm$   25 &   61 $\pm$    5 &  1.3 $\pm$  0.4 &   18 $\pm$    2 &   13 $\pm$    1 &  0.8 $\pm$  0.3 &      \nodata  &    6 $\pm$    1 \\
  09 &  1.5 $\pm$  0.6 &   \nodata       &   26 $\pm$    3 &  292 $\pm$   28 &   78 $\pm$    7 &  3.4 $\pm$  1.1 &   32 $\pm$    3 &   20 $\pm$    2 &      \nodata  &      \nodata  &    7 $\pm$    1 \\
  10 &  2.3 $\pm$  0.3 &   \nodata       &   31 $\pm$    3 &  286 $\pm$   27 &   92 $\pm$    9 &  2.3 $\pm$  0.3 &   34 $\pm$    3 &   24 $\pm$    2 &  2.7 $\pm$  0.4 &  1.2 $\pm$  0.3 &   15 $\pm$    3 \\
  11 & 42.4 $\pm$  3.2 &   \nodata       &   60 $\pm$    5 &  291 $\pm$   21 &  180 $\pm$   13 &  2.9 $\pm$  0.6 &   97 $\pm$    7 &   85 $\pm$    6 &  4.2 $\pm$  0.7 & 15.9 $\pm$  1.6 &   12 $\pm$    2 \\
  12 &  2.5 $\pm$  0.2 &   \nodata       &   28 $\pm$    3 &  288 $\pm$   28 &   84 $\pm$    8 &  3.0 $\pm$  0.6 &   35 $\pm$    4 &   24 $\pm$    3 &  3.4 $\pm$  0.6 &      \nodata  &   13 $\pm$    3 \\
  14 &  5.7 $\pm$  0.4 & 0.80 $\pm$ 0.12 &   34 $\pm$    2 &  286 $\pm$   20 &   93 $\pm$    6 &  3.5 $\pm$  0.3 &   43 $\pm$    3 &   35 $\pm$    2 &  6.1 $\pm$  0.5 &  5.4 $\pm$  0.4 &   19 $\pm$    4 \\
  15 & 11.5 $\pm$  1.2 &   \nodata       &   21 $\pm$    2 &  286 $\pm$   21 &   55 $\pm$    4 &  2.7 $\pm$  0.9 &   60 $\pm$    5 &   36 $\pm$    3 &  6.1 $\pm$  0.9 &  9.1 $\pm$  1.2 &   14 $\pm$    3 \\

%% file: ngc1365_b.tex
  03 &  242 $\pm$   20 &     \nodata   &   25 $\pm$    2 &   \nodata       &   44 $\pm$    3 &      \nodata  &   20 $\pm$    1 &   71 $\pm$    4 &   \nodata       &   \nodata       & 11.0 $\pm$  1.0 \\
  04 &  193 $\pm$   13 &     \nodata   &        \nodata  &   \nodata       &   46 $\pm$    3 &      \nodata  &   25 $\pm$    3 &   67 $\pm$    5 &   \nodata       &   \nodata       & 11.2 $\pm$  1.4 \\
  05 &  169 $\pm$   11 &     \nodata   &   25 $\pm$    1 & 2.43 $\pm$ 0.40 &   46 $\pm$    2 &      \nodata  &    9 $\pm$    1 &   25 $\pm$    1 & 2.32 $\pm$ 0.24 & 0.45 $\pm$ 0.18 &  9.4 $\pm$  0.6 \\
  06 &  164 $\pm$   11 &     \nodata   &   25 $\pm$    2 &   \nodata       &   46 $\pm$    3 &      \nodata  &   12 $\pm$    2 &   32 $\pm$    3 &   \nodata       &   \nodata       & 10.4 $\pm$  0.9 \\
  07 &  320 $\pm$   21 &     \nodata   &   25 $\pm$    2 &   \nodata       &   46 $\pm$    3 &      \nodata  &   45 $\pm$    3 &  134 $\pm$    8 &   \nodata       &   \nodata       & 10.7 $\pm$  1.5 \\
  08 &  172 $\pm$   11 & 3.9 $\pm$ 0.5 &   26 $\pm$    1 &   \nodata       &   46 $\pm$    2 &  3.5 $\pm$  0.3 &   20 $\pm$    1 &   60 $\pm$    4 & 1.73 $\pm$ 0.28 & 0.66 $\pm$ 0.15 & 11.2 $\pm$  0.7 \\
  09 &   28 $\pm$    2 &     \nodata   &   30 $\pm$    1 &   \nodata       &   46 $\pm$    2 &      \nodata  &    3 $\pm$    1 &    6 $\pm$    1 & 2.25 $\pm$ 0.39 &   \nodata       &  7.7 $\pm$  0.6 \\
  10 &  135 $\pm$   11 &     \nodata   &   25 $\pm$    1 &   \nodata       &   49 $\pm$    3 &      \nodata  &    6 $\pm$    1 &   17 $\pm$    1 &   \nodata       &   \nodata       &  9.4 $\pm$  0.8 \\
  11 &  165 $\pm$   16 &     \nodata   &        \nodata  &   \nodata       &   46 $\pm$    5 &      \nodata  &   11 $\pm$    3 &   26 $\pm$    4 &   \nodata       &   \nodata       & 10.0 $\pm$  1.2 \\
  12 &  175 $\pm$   14 &     \nodata   &   24 $\pm$    2 &   \nodata       &   47 $\pm$    3 &      \nodata  &   11 $\pm$    1 &   30 $\pm$    2 & 3.40 $\pm$ 0.68 &   \nodata       &  9.6 $\pm$  0.8 \\
  13 &  209 $\pm$   22 & 3.3 $\pm$ 1.4 &   26 $\pm$    2 &   \nodata       &   49 $\pm$    3 &      \nodata  &   15 $\pm$    1 &   46 $\pm$    3 &   \nodata       &   \nodata       & 11.8 $\pm$  1.2 \\
  14 &  209 $\pm$   13 & 4.4 $\pm$ 1.2 &   25 $\pm$    1 &   \nodata       &   46 $\pm$    2 &  3.4 $\pm$  0.5 &   13 $\pm$    1 &   39 $\pm$    2 & 1.90 $\pm$ 0.36 &   \nodata       &  9.8 $\pm$  0.7 \\
  15 &  221 $\pm$   18 & 2.7 $\pm$ 0.3 &   26 $\pm$    1 & 1.29 $\pm$ 0.18 &   53 $\pm$    3 &  3.7 $\pm$  0.3 &   26 $\pm$    2 &   74 $\pm$    4 & 1.52 $\pm$ 0.18 & 0.92 $\pm$ 0.11 & 12.7 $\pm$  1.0 \\
  16 &  244 $\pm$   16 & 6.5 $\pm$ 1.0 &   25 $\pm$    1 &   \nodata       &   48 $\pm$    3 &  5.1 $\pm$  0.8 &   34 $\pm$    2 &   93 $\pm$    6 & 2.14 $\pm$ 0.62 &   \nodata       & 12.4 $\pm$  0.9 \\
  17 &  360 $\pm$   22 & 8.3 $\pm$ 1.5 &   25 $\pm$    1 & 3.05 $\pm$ 0.63 &   46 $\pm$    3 &  3.8 $\pm$  1.1 &   35 $\pm$    2 &  100 $\pm$    6 & 2.90 $\pm$ 1.38 &   \nodata       & 10.7 $\pm$  0.8 \\

%% file: ngc1365_r.tex
  03 &  2.7 $\pm$  0.5 &   \nodata       &   28 $\pm$    3 &  287 $\pm$   27 &   85 $\pm$    8 &  3.6 $\pm$  0.6 &   27 $\pm$    3 &   20 $\pm$    2 &  6.6 $\pm$  0.9 &  2.4 $\pm$  0.8 &   30 $\pm$    6 \\
  04 &  3.2 $\pm$  1.0 &   \nodata       &   32 $\pm$    3 &  285 $\pm$   20 &   93 $\pm$    7 &  3.2 $\pm$  1.0 &   41 $\pm$    3 &   34 $\pm$    3 &  4.0 $\pm$  1.0 &      \nodata  &   16 $\pm$    3 \\
  05 &  2.4 $\pm$  0.2 & 0.32 $\pm$ 0.11 &   34 $\pm$    2 &  286 $\pm$   20 &  101 $\pm$    7 &  2.2 $\pm$  0.2 &   38 $\pm$    3 &   28 $\pm$    2 &  2.3 $\pm$  0.2 &  1.6 $\pm$  0.2 &   16 $\pm$    3 \\
  06 &  2.5 $\pm$  0.5 &   \nodata       &   31 $\pm$    2 &  287 $\pm$   20 &   91 $\pm$    7 &  1.9 $\pm$  0.4 &   39 $\pm$    3 &   28 $\pm$    2 &  1.6 $\pm$  0.4 &      \nodata  &   13 $\pm$    3 \\
  07 &  8.7 $\pm$  1.0 &   \nodata       &   27 $\pm$    2 &  285 $\pm$   20 &   81 $\pm$    6 &  2.2 $\pm$  0.7 &   58 $\pm$    4 &   39 $\pm$    3 &  4.3 $\pm$  0.8 &      \nodata  &   18 $\pm$    4 \\
  08 &  1.8 $\pm$  0.2 & 0.63 $\pm$ 0.13 &   29 $\pm$    2 &  286 $\pm$   20 &   86 $\pm$    6 &  2.8 $\pm$  0.2 &   30 $\pm$    2 &   22 $\pm$    2 &  4.1 $\pm$  0.3 &  2.0 $\pm$  0.3 &   19 $\pm$    4 \\
  09 &  1.4 $\pm$  0.2 &   \nodata       &   40 $\pm$    4 &  286 $\pm$   26 &  106 $\pm$   10 &  1.7 $\pm$  0.2 &   28 $\pm$    3 &   23 $\pm$    2 &  0.7 $\pm$  0.2 &      \nodata  &   12 $\pm$    2 \\
  10 &  0.8 $\pm$  0.2 &   \nodata       &   31 $\pm$    3 &  286 $\pm$   27 &   91 $\pm$    9 &  2.3 $\pm$  0.3 &   27 $\pm$    3 &   19 $\pm$    2 &  2.5 $\pm$  0.4 &      \nodata  &   13 $\pm$    3 \\
  11 &  5.1 $\pm$  0.8 &   \nodata       &   36 $\pm$    4 &  287 $\pm$   29 &  111 $\pm$   11 &  2.6 $\pm$  0.7 &   41 $\pm$    4 &   30 $\pm$    3 &  2.6 $\pm$  0.5 &      \nodata  &    9 $\pm$    2 \\
  12 &  3.6 $\pm$  0.5 &   \nodata       &   35 $\pm$    3 &  287 $\pm$   27 &  103 $\pm$   10 &  2.7 $\pm$  0.4 &   44 $\pm$    4 &   32 $\pm$    3 &  2.6 $\pm$  0.4 &  1.6 $\pm$  0.3 &   14 $\pm$    3 \\
  13 &  2.8 $\pm$  0.5 &   \nodata       &   33 $\pm$    4 &  306 $\pm$   38 &  101 $\pm$   13 &  2.9 $\pm$  0.5 &   38 $\pm$    5 &   27 $\pm$    4 &  4.1 $\pm$  0.7 &      \nodata  &   16 $\pm$    3 \\
  14 &  3.1 $\pm$  0.3 & 0.38 $\pm$ 0.16 &   31 $\pm$    2 &  287 $\pm$   20 &   91 $\pm$    6 &  2.6 $\pm$  0.3 &   36 $\pm$    3 &   25 $\pm$    2 &  3.1 $\pm$  0.3 &  1.3 $\pm$  0.2 &   15 $\pm$    3 \\
  15 &  2.2 $\pm$  0.2 & 1.00 $\pm$ 0.12 &   41 $\pm$    4 &  286 $\pm$   27 &  128 $\pm$   12 &  3.9 $\pm$  0.4 &   32 $\pm$    3 &   24 $\pm$    2 &  7.4 $\pm$  0.8 &  3.1 $\pm$  0.4 &   33 $\pm$    7 \\
  16 &  1.3 $\pm$  0.3 & 0.78 $\pm$ 0.28 &   24 $\pm$    2 &  286 $\pm$   20 &   67 $\pm$    5 &  3.2 $\pm$  0.4 &   25 $\pm$    2 &   18 $\pm$    1 &  4.7 $\pm$  0.5 &  4.0 $\pm$  0.6 &   19 $\pm$    4 \\
  17 &  4.9 $\pm$  0.5 &   \nodata       &   27 $\pm$    2 &  286 $\pm$   18 &   88 $\pm$    6 &  3.2 $\pm$  0.5 &   56 $\pm$    4 &   40 $\pm$    3 &  5.2 $\pm$  0.5 &  5.6 $\pm$  0.9 &   20 $\pm$    4 \\

%% file: ngc2903_b.tex
  02 &  213 $\pm$   15 &     \nodata   &   24 $\pm$    2 &   \nodata       &   46 $\pm$    3 &      \nodata  &   17 $\pm$    2 &   56 $\pm$    4 &   \nodata       &   \nodata       & 10.0 $\pm$  1.6 \\
  03 &  202 $\pm$   13 &     \nodata   &   26 $\pm$    1 &   \nodata       &   46 $\pm$    2 &      \nodata  &   32 $\pm$    2 &   92 $\pm$    6 &   \nodata       &   \nodata       & 12.1 $\pm$  1.2 \\
  04 &  233 $\pm$   20 &     \nodata   &   25 $\pm$    3 &   \nodata       &   46 $\pm$    3 &      \nodata  &   21 $\pm$    3 &   48 $\pm$    4 &   \nodata       &   \nodata       & 13.3 $\pm$  2.2 \\
  05 &  233 $\pm$   19 &     \nodata   &   25 $\pm$    1 &   \nodata       &   48 $\pm$    3 &  3.2 $\pm$  0.4 &   16 $\pm$    1 &   47 $\pm$    3 &   \nodata       &   \nodata       &  8.8 $\pm$  0.7 \\
  06 &  196 $\pm$   14 &     \nodata   &   25 $\pm$    2 &   \nodata       &   46 $\pm$    3 &      \nodata  &    9 $\pm$    2 &   25 $\pm$    2 &   \nodata       &   \nodata       & 11.2 $\pm$  1.9 \\
  07 &   48 $\pm$    4 &     \nodata   &   29 $\pm$    2 &   \nodata       &   46 $\pm$    2 &      \nodata  &    1 $\pm$    1 &    7 $\pm$    1 & 2.73 $\pm$ 0.30 &   \nodata       &  6.6 $\pm$  0.5 \\
  08 &   36 $\pm$    3 &     \nodata   &   29 $\pm$    1 &   \nodata       &   46 $\pm$    2 &      \nodata  &    1 $\pm$    1 &    3 $\pm$    1 & 1.86 $\pm$ 0.32 &   \nodata       &  5.8 $\pm$  0.5 \\
  12 &  165 $\pm$   10 &     \nodata   &   24 $\pm$    1 &   \nodata       &   46 $\pm$    2 &  4.3 $\pm$  0.4 &   13 $\pm$    1 &   42 $\pm$    3 & 1.27 $\pm$ 0.22 &   \nodata       &  9.0 $\pm$  0.6 \\
  14 &  212 $\pm$   23 &     \nodata   &   25 $\pm$    3 &   \nodata       &   43 $\pm$    4 &      \nodata  &   13 $\pm$    2 &   31 $\pm$    3 &   \nodata       &   \nodata       & 11.1 $\pm$  2.2 \\

%% file: ngc2903_r.tex
  02 &      \nodata  &   \nodata       &   31 $\pm$    3 &  283 $\pm$   20 &   91 $\pm$    7 &  4.4 $\pm$  1.5 &   48 $\pm$    4 &   33 $\pm$    3 &  5.6 $\pm$  1.4 &      \nodata  &   18 $\pm$    4 \\
  03 &      \nodata  &   \nodata       &   27 $\pm$    2 &  286 $\pm$   20 &   82 $\pm$    6 &      \nodata  &   35 $\pm$    3 &   25 $\pm$    2 &  5.7 $\pm$  0.9 &      \nodata  &   29 $\pm$    6 \\
  04 &      \nodata  &   \nodata       &   27 $\pm$    3 &  285 $\pm$   28 &   83 $\pm$    8 &  2.0 $\pm$  1.1 &   34 $\pm$    4 &   24 $\pm$    3 &  4.9 $\pm$  1.5 &      \nodata  &    9 $\pm$    2 \\
  05 &      \nodata  &   \nodata       &   32 $\pm$    3 &  287 $\pm$   27 &   96 $\pm$    9 &  2.8 $\pm$  0.4 &   43 $\pm$    4 &   30 $\pm$    3 &  4.8 $\pm$  0.6 &      \nodata  &   20 $\pm$    4 \\
  06 &      \nodata  &   \nodata       &   39 $\pm$    4 &  287 $\pm$   22 &  121 $\pm$    9 &  2.3 $\pm$  1.6 &   39 $\pm$    3 &   27 $\pm$    3 &  4.7 $\pm$  1.6 &  4.1 $\pm$  0.5 &   31 $\pm$    6 \\
  07 &      \nodata  &   \nodata       &   31 $\pm$    3 &  286 $\pm$   27 &   92 $\pm$    9 &  1.7 $\pm$  0.2 &   26 $\pm$    3 &   19 $\pm$    2 &  0.7 $\pm$  0.1 &      \nodata  &    9 $\pm$    2 \\
  08 &      \nodata  &   \nodata       &   30 $\pm$    3 &  287 $\pm$   27 &   90 $\pm$    8 &  1.7 $\pm$  0.2 &   22 $\pm$    2 &   17 $\pm$    2 &  0.9 $\pm$  0.2 &      \nodata  &   11 $\pm$    2 \\
  12 &      \nodata  &   \nodata       &   26 $\pm$    2 &  286 $\pm$   20 &   80 $\pm$    6 &  2.7 $\pm$  0.3 &   31 $\pm$    2 &   22 $\pm$    2 &  3.5 $\pm$  0.3 &      \nodata  &   18 $\pm$    4 \\
  14 &      \nodata  &   \nodata       &   33 $\pm$    4 &  264 $\pm$   34 &  105 $\pm$   14 &  3.7 $\pm$  0.8 &   34 $\pm$    5 &   24 $\pm$    3 &  4.6 $\pm$  1.0 &      \nodata  &   23 $\pm$    5 \\

%% file: ngc2997_b.tex
  03 &  235 $\pm$   19 &     \nodata   &   24 $\pm$    2 &   \nodata       &   44 $\pm$    3 &  3.7 $\pm$  0.7 &   21 $\pm$    1 &   61 $\pm$    4 &   \nodata       &   \nodata       &  7.7 $\pm$  0.7 \\
  04 &  383 $\pm$   31 &10.2 $\pm$ 1.0 &   26 $\pm$    1 &   \nodata       &   46 $\pm$    3 &  3.8 $\pm$  0.4 &   51 $\pm$    3 &  153 $\pm$    9 & 3.02 $\pm$ 0.50 &   \nodata       &  6.2 $\pm$  0.5 \\
  05 &  213 $\pm$   17 & 3.3 $\pm$ 0.4 &   25 $\pm$    1 &   \nodata       &   46 $\pm$    2 &  3.9 $\pm$  0.4 &   28 $\pm$    2 &   82 $\pm$    5 & 2.07 $\pm$ 0.31 & 0.46 $\pm$ 0.16 &  7.6 $\pm$  0.6 \\
  06 &  215 $\pm$   13 & 2.7 $\pm$ 0.3 &   29 $\pm$    1 & 1.65 $\pm$ 0.27 &   46 $\pm$    2 &  3.8 $\pm$  0.3 &   28 $\pm$    2 &   81 $\pm$    5 & 0.98 $\pm$ 0.16 & 0.58 $\pm$ 0.10 & 12.0 $\pm$  0.8 \\
  07 &  203 $\pm$   16 & 3.1 $\pm$ 0.7 &   25 $\pm$    1 &   \nodata       &   49 $\pm$    3 &  2.7 $\pm$  0.3 &   15 $\pm$    1 &   44 $\pm$    3 & 2.56 $\pm$ 0.30 & 0.64 $\pm$ 0.16 &  8.8 $\pm$  0.7 \\
  08 &  120 $\pm$   10 &     \nodata   &   23 $\pm$    2 &   \nodata       &   46 $\pm$    3 &      \nodata  &    7 $\pm$    1 &   19 $\pm$    1 &   \nodata       &   \nodata       &  9.0 $\pm$  0.9 \\
  09 &  142 $\pm$    9 &     \nodata   &   22 $\pm$    1 &   \nodata       &   46 $\pm$    3 &      \nodata  &   12 $\pm$    1 &   33 $\pm$    2 & 2.92 $\pm$ 0.60 &   \nodata       & 10.0 $\pm$  0.8 \\
  10 &  116 $\pm$    7 &     \nodata   &   26 $\pm$    1 &   \nodata       &   46 $\pm$    2 &      \nodata  &    8 $\pm$    1 &   22 $\pm$    1 &   \nodata       &   \nodata       &  9.2 $\pm$  0.7 \\
  11 &  171 $\pm$   10 &     \nodata   &   25 $\pm$    1 &   \nodata       &   46 $\pm$    2 &      \nodata  &   13 $\pm$    1 &   41 $\pm$    3 &   \nodata       &   \nodata       &  8.1 $\pm$  0.7 \\
  12 &   80 $\pm$    6 &     \nodata   &   27 $\pm$    2 &   \nodata       &   46 $\pm$    3 &      \nodata  &    4 $\pm$    1 &   11 $\pm$    1 &   \nodata       &   \nodata       &  7.0 $\pm$  0.6 \\
  13 &  118 $\pm$    7 &     \nodata   &   28 $\pm$    1 &   \nodata       &   46 $\pm$    2 &  3.0 $\pm$  0.6 &    6 $\pm$    1 &   19 $\pm$    1 & 2.34 $\pm$ 0.51 &   \nodata       &  8.2 $\pm$  0.6 \\
  14 &  177 $\pm$   11 &     \nodata   &   29 $\pm$    1 &   \nodata       &   46 $\pm$    2 &      \nodata  &   15 $\pm$    1 &   44 $\pm$    3 & 2.46 $\pm$ 0.56 &   \nodata       & 10.3 $\pm$  0.8 \\
  15 &  210 $\pm$   13 &     \nodata   &   30 $\pm$    1 &   \nodata       &   46 $\pm$    2 &      \nodata  &   18 $\pm$    1 &   54 $\pm$    3 & 1.90 $\pm$ 0.45 &   \nodata       &  8.3 $\pm$  0.6 \\
  16 &  312 $\pm$   18 & 9.8 $\pm$ 1.5 &   25 $\pm$    2 &   \nodata       &   46 $\pm$    3 &      \nodata  &   20 $\pm$    1 &   56 $\pm$    3 &   \nodata       &   \nodata       & 11.1 $\pm$  0.9 \\

%% file: ngc2997_r.tex
  03 &  2.1 $\pm$  0.3 &   \nodata       &   23 $\pm$    2 &  266 $\pm$   25 &   69 $\pm$    7 &  2.3 $\pm$  0.4 &   32 $\pm$    3 &   22 $\pm$    2 &  4.3 $\pm$  0.6 &      \nodata  &   24 $\pm$    5 \\
  04 &  5.6 $\pm$  0.5 & 0.96 $\pm$ 0.18 &   22 $\pm$    2 &  286 $\pm$   27 &   58 $\pm$    6 &  2.7 $\pm$  0.3 &   37 $\pm$    4 &   26 $\pm$    3 &  7.6 $\pm$  0.9 &  6.6 $\pm$  0.8 &   26 $\pm$    5 \\
  05 &  3.8 $\pm$  0.4 & 0.54 $\pm$ 0.11 &   29 $\pm$    3 &  287 $\pm$   27 &   87 $\pm$    8 &  3.0 $\pm$  0.3 &   38 $\pm$    4 &   27 $\pm$    3 &  5.9 $\pm$  0.7 &  3.4 $\pm$  0.4 &   25 $\pm$    5 \\
  06 &  1.7 $\pm$  0.2 & 0.96 $\pm$ 0.12 &   29 $\pm$    2 &  287 $\pm$   20 &   84 $\pm$    6 &  3.3 $\pm$  0.3 &   21 $\pm$    1 &   15 $\pm$    1 &  6.6 $\pm$  0.5 &  3.2 $\pm$  0.3 &   30 $\pm$    6 \\
  07 &  3.1 $\pm$  0.3 & 0.61 $\pm$ 0.13 &   37 $\pm$    4 &  287 $\pm$   27 &   97 $\pm$    9 &  2.8 $\pm$  0.3 &   32 $\pm$    3 &   24 $\pm$    2 &  4.2 $\pm$  0.5 &  2.5 $\pm$  0.3 &   23 $\pm$    5 \\
  08 &  1.6 $\pm$  0.4 &   \nodata       &   31 $\pm$    3 &  286 $\pm$   27 &   96 $\pm$    9 &  2.1 $\pm$  0.4 &   31 $\pm$    3 &   22 $\pm$    2 &  2.5 $\pm$  0.4 &      \nodata  &   11 $\pm$    2 \\
  09 &  3.8 $\pm$  0.5 &   \nodata       &   32 $\pm$    2 &  286 $\pm$   20 &   96 $\pm$    7 &  2.4 $\pm$  0.4 &   45 $\pm$    3 &   32 $\pm$    2 &  2.8 $\pm$  0.4 &      \nodata  &   13 $\pm$    3 \\
  10 &  2.4 $\pm$  0.4 &   \nodata       &   30 $\pm$    2 &  286 $\pm$   20 &   89 $\pm$    6 &  2.3 $\pm$  0.4 &   33 $\pm$    2 &   24 $\pm$    2 &  1.9 $\pm$  0.3 &      \nodata  &   18 $\pm$    4 \\
  11 &  9.4 $\pm$  0.8 &   \nodata       &   36 $\pm$    2 &  287 $\pm$   19 &  105 $\pm$    7 &  2.6 $\pm$  0.4 &   51 $\pm$    4 &   38 $\pm$    3 &  1.9 $\pm$  0.4 &  2.7 $\pm$  0.7 &   20 $\pm$    4 \\
  12 &  2.5 $\pm$  0.4 &   \nodata       &   33 $\pm$    3 &  286 $\pm$   22 &  100 $\pm$    8 &  1.9 $\pm$  0.3 &   30 $\pm$    2 &   21 $\pm$    2 &  1.7 $\pm$  0.3 &      \nodata  &   16 $\pm$    3 \\
  13 &  2.5 $\pm$  0.2 & 0.28 $\pm$ 0.09 &   35 $\pm$    2 &  287 $\pm$   20 &  106 $\pm$    7 &  2.4 $\pm$  0.2 &   29 $\pm$    2 &   22 $\pm$    2 &  3.1 $\pm$  0.3 &  1.7 $\pm$  0.2 &   26 $\pm$    5 \\
  14 &  3.8 $\pm$  0.5 &   \nodata       &   32 $\pm$    2 &  287 $\pm$   20 &   97 $\pm$    7 &  2.5 $\pm$  0.4 &   37 $\pm$    3 &   26 $\pm$    2 &  3.4 $\pm$  0.4 &  2.4 $\pm$  0.7 &   25 $\pm$    5 \\
  15 &  2.5 $\pm$  0.3 &   \nodata       &   28 $\pm$    2 &  286 $\pm$   20 &   83 $\pm$    6 &  2.7 $\pm$  0.3 &   34 $\pm$    2 &   25 $\pm$    2 &  4.9 $\pm$  0.4 &      \nodata  &   29 $\pm$    6 \\
  16 &  3.6 $\pm$  0.6 &   \nodata       &   30 $\pm$    2 &  286 $\pm$   17 &   89 $\pm$    5 &  2.6 $\pm$  0.6 &   51 $\pm$    3 &   35 $\pm$    2 &  4.5 $\pm$  0.6 &      \nodata  &   20 $\pm$    4 \\

%% file: ngc5236_b.tex
  02 &  162 $\pm$   11 &     \nodata   &   24 $\pm$    2 &   \nodata       &   46 $\pm$    3 &      \nodata  &    6 $\pm$    1 &   16 $\pm$    1 &   \nodata       &   \nodata       &  8.0 $\pm$  0.7 \\
  03 &  208 $\pm$   17 & 6.9 $\pm$ 0.6 &   25 $\pm$    1 &   \nodata       &   49 $\pm$    3 &  3.7 $\pm$  0.3 &   39 $\pm$    2 &  118 $\pm$    7 & 0.96 $\pm$ 0.18 & 0.74 $\pm$ 0.17 & 10.8 $\pm$  0.8 \\
  04 &   72 $\pm$    4 &     \nodata   &   24 $\pm$    1 &   \nodata       &   46 $\pm$    2 &      \nodata  &        \nodata  &    4 $\pm$    1 & 1.70 $\pm$ 1.14 &   \nodata       &  7.7 $\pm$  0.5 \\
  05 &   68 $\pm$    5 &     \nodata   &   25 $\pm$    2 &   \nodata       &   49 $\pm$    3 &  3.4 $\pm$  0.7 &    6 $\pm$    1 &   17 $\pm$    1 &   \nodata       & 0.94 $\pm$ 0.14 &  8.5 $\pm$  0.7 \\
  06 &  123 $\pm$    8 &     \nodata   &   25 $\pm$    1 &   \nodata       &   46 $\pm$    2 &  2.5 $\pm$  0.5 &    7 $\pm$    1 &   21 $\pm$    1 & 1.38 $\pm$ 0.56 & 0.81 $\pm$ 0.33 &  9.1 $\pm$  0.6 \\
  07 &   91 $\pm$    7 &     \nodata   &   25 $\pm$    1 &   \nodata       &   46 $\pm$    2 &      \nodata  &    3 $\pm$    1 &   10 $\pm$    1 &   \nodata       &   \nodata       &  9.0 $\pm$  0.7 \\
  08 &   54 $\pm$    4 &     \nodata   &   30 $\pm$    1 &   \nodata       &   46 $\pm$    2 &      \nodata  &    2 $\pm$    1 &    5 $\pm$    1 & 1.64 $\pm$ 0.38 &   \nodata       &  7.0 $\pm$  0.5 \\
  09 &   49 $\pm$    4 &     \nodata   &   25 $\pm$    1 &   \nodata       &   48 $\pm$    3 &  3.0 $\pm$  0.4 &    2 $\pm$    1 &    5 $\pm$    1 & 1.93 $\pm$ 0.32 &   \nodata       &  8.8 $\pm$  0.6 \\
  10 &   76 $\pm$    6 &     \nodata   &   25 $\pm$    1 &   \nodata       &   46 $\pm$    2 &      \nodata  &    2 $\pm$    1 &    6 $\pm$    1 & 1.87 $\pm$ 0.30 &   \nodata       &  8.8 $\pm$  0.7 \\
  11 &   47 $\pm$    3 &     \nodata   &   25 $\pm$    1 &   \nodata       &   46 $\pm$    2 &  3.3 $\pm$  0.2 &    3 $\pm$    1 &    9 $\pm$    1 & 1.53 $\pm$ 0.16 & 0.40 $\pm$ 0.05 & 10.7 $\pm$  0.6 \\
  12 &  107 $\pm$    8 &     \nodata   &   25 $\pm$    1 &   \nodata       &   49 $\pm$    3 &  3.1 $\pm$  0.3 &    2 $\pm$    1 &    5 $\pm$    1 & 0.89 $\pm$ 0.26 &   \nodata       &  8.0 $\pm$  0.6 \\
  13 &  111 $\pm$    9 &     \nodata   &   25 $\pm$    1 &   \nodata       &   46 $\pm$    2 &  2.9 $\pm$  0.3 &    6 $\pm$    1 &   19 $\pm$    1 & 1.28 $\pm$ 0.24 &   \nodata       &      \nodata  \\
  14 &   74 $\pm$    5 &     \nodata   &   25 $\pm$    1 &   \nodata       &   46 $\pm$    2 &      \nodata  &    3 $\pm$    1 &    9 $\pm$    1 & 1.89 $\pm$ 0.33 & 0.42 $\pm$ 0.17 &  8.2 $\pm$  0.6 \\
  15 &   89 $\pm$    6 &     \nodata   &        \nodata  &   \nodata       &   46 $\pm$    2 &      \nodata  &    5 $\pm$    1 &   14 $\pm$    1 &   \nodata       &   \nodata       &  7.4 $\pm$  0.7 \\
  16 &  144 $\pm$    9 &     \nodata   &   25 $\pm$    1 &   \nodata       &   46 $\pm$    2 &  2.7 $\pm$  0.3 &    8 $\pm$    1 &   23 $\pm$    1 & 1.78 $\pm$ 0.29 & 0.60 $\pm$ 0.11 &  8.3 $\pm$  0.5 \\
  17 &  129 $\pm$   13 &     \nodata   &   21 $\pm$    1 &   \nodata       &   46 $\pm$    3 &  2.7 $\pm$  0.4 &    4 $\pm$    1 &   12 $\pm$    1 & 2.25 $\pm$ 0.42 &   \nodata       &  8.0 $\pm$  0.8 \\
  18 &  155 $\pm$   13 &     \nodata   &   25 $\pm$    2 &   \nodata       &   46 $\pm$    3 &      \nodata  &    3 $\pm$    1 &    9 $\pm$    1 & 3.53 $\pm$ 1.28 &   \nodata       & 10.5 $\pm$  1.0 \\
  19 &  166 $\pm$   14 &     \nodata   &   25 $\pm$    4 &   \nodata       &   46 $\pm$    5 &      \nodata  &   24 $\pm$    5 &   55 $\pm$    6 &   \nodata       &   \nodata       &  9.5 $\pm$  1.3 \\

%% file: ngc5236_r.tex
  02 &      \nodata  &   \nodata       &   35 $\pm$    2 &  287 $\pm$   20 &  108 $\pm$    7 &  2.0 $\pm$  0.3 &   31 $\pm$    2 &   22 $\pm$    2 &  2.5 $\pm$  0.3 &      \nodata  &   18 $\pm$    4 \\
  03 &  1.0 $\pm$  0.1 & 0.72 $\pm$ 0.13 &   32 $\pm$    3 &  306 $\pm$   29 &   98 $\pm$    9 &  3.4 $\pm$  0.4 &   25 $\pm$    2 &   18 $\pm$    2 &  8.4 $\pm$  0.9 &  2.9 $\pm$  0.4 &   26 $\pm$    5 \\
  04 &      \nodata  &   \nodata       &   33 $\pm$    2 &  287 $\pm$   20 &   90 $\pm$    6 &  1.8 $\pm$  0.2 &   28 $\pm$    2 &   20 $\pm$    1 &  0.9 $\pm$  0.2 &      \nodata  &   10 $\pm$    2 \\
  05 &      \nodata  &   \nodata       &   34 $\pm$    3 &  286 $\pm$   27 &  100 $\pm$    9 &  2.5 $\pm$  0.3 &   18 $\pm$    2 &   13 $\pm$    1 &  3.2 $\pm$  0.4 &  0.9 $\pm$  0.2 &   17 $\pm$    3 \\
  06 &  0.9 $\pm$  0.2 & 0.50 $\pm$ 0.17 &   41 $\pm$    3 &  286 $\pm$   20 &  124 $\pm$    8 &  2.4 $\pm$  0.3 &   28 $\pm$    2 &   21 $\pm$    2 &  2.9 $\pm$  0.4 &  1.2 $\pm$  0.3 &   24 $\pm$    5 \\
  07 &      \nodata  &   \nodata       &   34 $\pm$    3 &  287 $\pm$   26 &  102 $\pm$    9 &  1.6 $\pm$  0.3 &   31 $\pm$    3 &   22 $\pm$    2 &  1.5 $\pm$  0.4 &      \nodata  &   15 $\pm$    3 \\
  08 &      \nodata  &   \nodata       &   29 $\pm$    2 &  286 $\pm$   23 &   83 $\pm$    7 &  1.5 $\pm$  0.2 &   22 $\pm$    2 &   16 $\pm$    1 &  0.6 $\pm$  0.2 &      \nodata  &    8 $\pm$    2 \\
  09 &  1.0 $\pm$  0.1 &   \nodata       &   33 $\pm$    3 &  287 $\pm$   27 &  100 $\pm$    9 &  2.3 $\pm$  0.2 &   22 $\pm$    2 &   18 $\pm$    2 &  1.3 $\pm$  0.2 &  0.7 $\pm$  0.1 &   11 $\pm$    2 \\
  10 &  1.0 $\pm$  0.2 &   \nodata       &   34 $\pm$    3 &  287 $\pm$   27 &   98 $\pm$    9 &  2.0 $\pm$  0.2 &   27 $\pm$    3 &   20 $\pm$    2 &  1.1 $\pm$  0.2 &      \nodata  &   11 $\pm$    2 \\
  11 &  1.1 $\pm$  0.1 & 0.23 $\pm$ 0.05 &   55 $\pm$    4 &  288 $\pm$   19 &  162 $\pm$   11 &  3.4 $\pm$  0.2 &   28 $\pm$    2 &   31 $\pm$    2 &  2.7 $\pm$  0.2 &  1.2 $\pm$  0.1 &   41 $\pm$    8 \\
  12 &  0.9 $\pm$  0.1 &   \nodata       &   37 $\pm$    3 &  287 $\pm$   27 &  112 $\pm$   11 &  2.3 $\pm$  0.3 &   26 $\pm$    3 &   19 $\pm$    2 &  2.0 $\pm$  0.2 &      \nodata  &   15 $\pm$    3 \\
  13 &      \nodata  &   \nodata       &        \nodata  &        \nodata  &        \nodata  &      \nodata  &        \nodata  &        \nodata  &      \nodata  &      \nodata  &        \nodata  \\
  14 &  1.1 $\pm$  0.2 &   \nodata       &   37 $\pm$    3 &  288 $\pm$   24 &  109 $\pm$    9 &  2.2 $\pm$  0.2 &   26 $\pm$    2 &   20 $\pm$    2 &  1.6 $\pm$  0.2 &  0.6 $\pm$  0.1 &   14 $\pm$    3 \\
  15 &      \nodata  &   \nodata       &   28 $\pm$    2 &  286 $\pm$   23 &   82 $\pm$    7 &  3.9 $\pm$  0.6 &   25 $\pm$    2 &   17 $\pm$    2 &  2.0 $\pm$  0.5 &      \nodata  &   15 $\pm$    3 \\
  16 &  1.8 $\pm$  0.2 & 0.35 $\pm$ 0.10 &   44 $\pm$    3 &  287 $\pm$   20 &  135 $\pm$    9 &  2.4 $\pm$  0.2 &   28 $\pm$    2 &   23 $\pm$    2 &  4.5 $\pm$  0.3 &  2.4 $\pm$  0.2 &   37 $\pm$    7 \\
  17 &  1.7 $\pm$  0.3 &   \nodata       &   39 $\pm$    5 &  287 $\pm$   36 &  115 $\pm$   15 &  2.0 $\pm$  0.3 &   31 $\pm$    4 &   24 $\pm$    3 &  2.1 $\pm$  0.4 &      \nodata  &   15 $\pm$    3 \\
  18 &  1.4 $\pm$  0.4 &   \nodata       &   37 $\pm$    4 &  288 $\pm$   27 &  110 $\pm$   11 &  1.8 $\pm$  0.4 &   34 $\pm$    3 &   25 $\pm$    3 &  2.3 $\pm$  0.5 &      \nodata  &   13 $\pm$    3 \\
  19 &      \nodata  &   \nodata       &   32 $\pm$    3 &  286 $\pm$   23 &   99 $\pm$    8 &  3.0 $\pm$  0.9 &   34 $\pm$    3 &   27 $\pm$    3 &  2.9 $\pm$  1.0 &      \nodata  &   10 $\pm$    2 \\

%% file: te.tex
 & & & & \\[-6mm]
\multicolumn{5}{c}{\em NGC~1232}\\[1mm]
02\dotfill &   11200 $\pm$  1300 &     \nodata &      \nodata &      \nodata \\
04\dotfill &   10800 $\pm$  800 &      \nodata &  10900 $\pm$  700 &   10400 $\pm$ 1500\\
05\dotfill &    8600 $\pm$  300 &   7800 $\pm$  700 &   7200 $\pm$  400 &    7200 $\pm$  400\\
06\dotfill &    9400 $\pm$  700 &   8000 $\pm$ 1000 &      \nodata &       \nodata\\
07\dotfill &    7300 $\pm$  400 &   6400 $\pm$  700 &   6800 $\pm$  800 & \nodata\\
10\dotfill &    7400 $\pm$  600 &      \nodata &      \nodata &       \nodata\\
11\dotfill &   16400 $\pm$ 1500 &      \nodata &      \nodata &   14700 $\pm$ 2300\\
14\dotfill &    9000 $\pm$  400 &   8500 $\pm$  400 &   8339 $\pm$  500 &    8100 $\pm$  400\\
15\dotfill &   10900 $\pm$  800 &      \nodata &      \nodata &       \nodata\\[2mm]
\multicolumn{5}{c}{\em NGC~1365}\\[1mm]
03\dotfill &    7800 $\pm$  1000 &      \nodata &      \nodata &       \nodata\\
05\dotfill &    7600 $\pm$  400 &   7000 $\pm$  800 &   6600 $\pm$  700 &    6300 $\pm$  500\\
08\dotfill &    8200 $\pm$  600 &   8200 $\pm$  600 &   7700 $\pm$  500 &       \nodata\\
12\dotfill &    7400 $\pm$  600 &      \nodata &      \nodata &       \nodata\\
14\dotfill &    6500 $\pm$  400 &      \nodata &   7100 $\pm$ 1000 &       \nodata\\
15\dotfill &    8700 $\pm$  500 &   8000 $\pm$  300 &   7500 $\pm$  300 &    5100 $\pm$  300\\
16\dotfill &    9700 $\pm$  800 &      \nodata &   8200 $\pm$ 1100 &       \nodata\\
17\dotfill &    9500 $\pm$  800 &      \nodata &      \nodata &    6000 $\pm$  500\\[2mm]
\multicolumn{5}{c}{\em NGC~2903}\\[1mm]
06\dotfill &   11200 $\pm$  800 &      \nodata &      \nodata &       \nodata\\[2mm]
\multicolumn{5}{c}{\em NGC~2997}\\[1mm]
04\dotfill &   10000 $\pm$  600 &      \nodata &   8000 $\pm$  500 &       \nodata\\
05\dotfill &    9600 $\pm$  600 &   7300 $\pm$  700 &   6700 $\pm$  400 &       \nodata\\
06\dotfill &    9100 $\pm$  400 &   7900 $\pm$  500 &   7700 $\pm$  300 &    7000 $\pm$  600\\
07\dotfill &    8300 $\pm$  500 &   7700 $\pm$  600 &   7200 $\pm$  500 &       \nodata\\
11\dotfill &    9300 $\pm$ 1200 &      \nodata &      \nodata &       \nodata\\
13\dotfill &    8700 $\pm$  400 &      \nodata &   5600 $\pm$  400 &       \nodata\\
14\dotfill &    8800 $\pm$ 1200 &      \nodata &      \nodata &       \nodata\\[2mm]
\multicolumn{5}{c}{\em NGC~5236}\\[1mm]
03\dotfill &    8800 $\pm$  500 &   8100 $\pm$  600 &   7300 $\pm$  400 &       \nodata\\
05\dotfill &    $<$ 8700 $\pm$ 1000 &   8700 $\pm$  400 &      \nodata &       \nodata\\
06\dotfill &    7500 $\pm$  800 &   7700 $\pm$ 1000 &   6700 $\pm$  700 &       \nodata\\
09\dotfill &    8100 $\pm$  700 &      \nodata &      \nodata &       \nodata\\
11\dotfill &    8900 $\pm$  400 &   5900 $\pm$  200 &   4800 $\pm$  200 &       \nodata\\
14\dotfill &    7000 $\pm$  600 &   6600 $\pm$  800 &      \nodata &       \nodata\\
16\dotfill &    9000 $\pm$  400 &   6900 $\pm$  400 &   5400 $\pm$  300 &       \nodata\\

%% file: te_adopted.tex
 & & & \\[-6mm]
\multicolumn{4}{c}{\em NGC~1232}\\[1mm]
02\dotfill & 11200 $\pm$  1300 & 11400 $\pm$  1500 & 11700 $\pm$  1800 \\
04\dotfill & 10700 $\pm$   600 & 10900 $\pm$   700 & 11100 $\pm$   800 \\
05\dotfill &  7700 $\pm$   700 &  7200 $\pm$   400 &  6700 $\pm$   400 \\
06\dotfill &  8000 $\pm$  1000 &  7700 $\pm$  1200 &  7200 $\pm$  1500 \\
07\dotfill &  6400 $\pm$   700 &  6800 $\pm$   800 &  5500 $\pm$   700 \\
10\dotfill &  7400 $\pm$   600 &  6900 $\pm$   800 &  6300 $\pm$   900 \\
11\dotfill & 16400 $\pm$  1500 & 17600 $\pm$  1800 & 19100 $\pm$  2200 \\
14\dotfill &  8500 $\pm$   400 &  8300 $\pm$   500 &  7900 $\pm$   400 \\
15\dotfill & 10900 $\pm$   800 & 11100 $\pm$  1000 & 11300 $\pm$  1200 \\[2mm]
\multicolumn{4}{c}{\em NGC~1365}\\[1mm]
03\dotfill &  7800 $\pm$  1000 &  7400 $\pm$  1200 &  6800 $\pm$  1400 \\
05\dotfill &  6900 $\pm$   800 &  6600 $\pm$   700 &  5800 $\pm$   600 \\
08\dotfill &  8200 $\pm$   600 &  7700 $\pm$   500 &  7300 $\pm$   500 \\
12\dotfill &  7400 $\pm$   600 &  6900 $\pm$   700 &  6200 $\pm$   800 \\
14\dotfill &  7500 $\pm$   900 &  7100 $\pm$  1000 &  6500 $\pm$  1100 \\
15\dotfill &  8000 $\pm$   300 &  7500 $\pm$   300 &  7100 $\pm$   200 \\
16\dotfill &  8500 $\pm$  1000 &  8300 $\pm$  1100 &  7900 $\pm$  1200 \\
17\dotfill &  9500 $\pm$   800 &  9400 $\pm$   900 &  9300 $\pm$  1100 \\[2mm]
\multicolumn{4}{c}{\em NGC~2903}\\[1mm]
06\dotfill & 11200 $\pm$   800 & 11500 $\pm$  1000 & 11800 $\pm$  1200 \\[2mm]
\multicolumn{4}{c}{\em NGC~2997}\\[1mm]
04\dotfill &  8300 $\pm$   500 &  8000 $\pm$   500 &  7600 $\pm$   600 \\
05\dotfill &  7300 $\pm$   700 &  6700 $\pm$   400 &  6100 $\pm$   400 \\
06\dotfill &  7900 $\pm$   500 &  7700 $\pm$   300 &  7100 $\pm$   300 \\
07\dotfill &  7700 $\pm$   600 &  7200 $\pm$   500 &  6600 $\pm$   400 \\
11\dotfill &  9300 $\pm$  1200 &  9200 $\pm$  1400 &  9000 $\pm$  1700 \\
13\dotfill &  6300 $\pm$   400 &  5600 $\pm$   400 &  4700 $\pm$   500 \\
14\dotfill &  8800 $\pm$  1200 &  8600 $\pm$  1400 &  8300 $\pm$  1700 \\[2mm]
\multicolumn{4}{c}{\em NGC~5236}\\[1mm]
03\dotfill &  8100 $\pm$   600 &  7300 $\pm$   400 &  7000 $\pm$   400 \\
05\dotfill &  8700 $\pm$   400 &  8500 $\pm$   500 &  8200 $\pm$   500 \\
06\dotfill &  7700 $\pm$  1000 &  6700 $\pm$   700 &  6300 $\pm$   700 \\
09\dotfill &  8100 $\pm$   700 &  7800 $\pm$   800 &  7300 $\pm$   900 \\
11\dotfill &  5900 $\pm$   200 &  4800 $\pm$   200 &  4000 $\pm$   200 \\
14\dotfill &  6600 $\pm$   800 &  6000 $\pm$   900 &  5200 $\pm$   900 \\
16\dotfill &  6900 $\pm$   400 &  5400 $\pm$   400 &  5000 $\pm$   300 \\

%% file: abund.tex
 & & & \\[-6mm]
\multicolumn{4}{c}{\em NGC~1232}\\[1mm]
02  \phantom{*} &    8.07 $\pm$    0.19  &   $-$1.11  $\pm$   0.34 &   $-$1.48  $\pm$   0.27 \\
04 *  &    8.20 $\pm$    0.09  &   $-$1.20  $\pm$   0.14 &   $-$1.73  $\pm$   0.11 \\
05 *  &    8.72 $\pm$    0.17  &   $-$0.99  $\pm$   0.36 &   $-$1.68  $\pm$   0.22 \\
06 *  &    8.47 $\pm$    0.32  &   $-$0.84  $\pm$   0.60 &   $-$1.89  $\pm$   0.51 \\
07 *  &    8.90 $\pm$    0.31  &   $-$0.84  $\pm$   0.58 &   $-$1.82  $\pm$   0.49 \\
10 \phantom{*}  &    8.44 $\pm$    0.22  &   $-$0.69  $\pm$   0.37 &   $-$1.51  $\pm$   0.31 \\
11 \phantom{*}  &    7.43 $\pm$    0.12  &   $-$0.29  $\pm$   0.18 &   $-$1.06  $\pm$   0.15 \\
14 *  &    8.50 $\pm$    0.10  &   $-$0.93  $\pm$   0.15 &   $-$1.65  $\pm$   0.12  \\
15 \phantom{*}  &    8.16 $\pm$    0.13  &   $-$1.12  $\pm$   0.20 &    $-$1.61 $\pm$   0.16 \\[2mm]
\multicolumn{4}{c}{\em NGC~1365}\\[1mm]
03 \phantom{*}  &    8.62 $\pm$    0.33  &   $-$0.94  $\pm$   0.61 &   $-$1.56  $\pm$   0.54 \\
05 *  &    8.71 $\pm$    0.31  &   $-$0.84  $\pm$   0.60 &   $-$1.67  $\pm$   0.47 \\
08 *  &    8.38 $\pm$    0.15  &   $-$0.74  $\pm$   0.28 &   $-$1.52  $\pm$   0.20 \\
12 \phantom{*}  &    8.57 $\pm$    0.20  &   $-$0.77  $\pm$   0.33 &   $-$1.61  $\pm$   0.28 \\
14 *  &    8.58 $\pm$    0.29  &   $-$0.86  $\pm$   0.55 &   $-$1.67  $\pm$   0.46 \\
15 *  &    8.54 $\pm$    0.08  &   $-$0.71  $\pm$   0.13 &   $-$1.47  $\pm$   0.09 \\
16 *  &    8.44 $\pm$    0.25  &   $-$0.97  $\pm$   0.49 &   $-$1.66  $\pm$   0.38 \\
17 \phantom{*}  &    8.33 $\pm$    0.16  &   $-$0.93  $\pm$   0.27 &   $-$1.57  $\pm$   0.21 \\[2mm]
\multicolumn{4}{c}{\em NGC~2903}\\[1mm]
06 \phantom{*}  &    7.70 $\pm$    0.13  &   $-$0.41  $\pm$   0.19 &   $-$1.01  $\pm$   0.16 \\[2mm]
\multicolumn{4}{c}{\em NGC~2997}\\[1mm]
04 *  &    8.71 $\pm$    0.12  &   $-$1.25  $\pm$   0.20 &   $-$1.75  $\pm$   0.15 \\
05 *  &    8.82 $\pm$    0.22  &   $-$0.97  $\pm$   0.48 &   $-$1.68  $\pm$   0.30 \\
06 *  &    8.57 $\pm$    0.11  &   $-$0.90  $\pm$   0.20 &   $-$1.58  $\pm$   0.14 \\
07 *  &    8.56 $\pm$    0.17  &   $-$0.81  $\pm$   0.31 &   $-$1.54  $\pm$   0.23 \\
11 \phantom{*}  &    8.04 $\pm$    0.26  &   $-$0.54  $\pm$   0.48 &   $-$1.24  $\pm$   0.39 \\
13 *  &    8.92 $\pm$    0.19  &   $-$0.81  $\pm$   0.30 &   $-$1.57  $\pm$   0.27 \\
14 \phantom{*}  &    8.17 $\pm$    0.30  &   $-$0.64  $\pm$   0.58 &   $-$1.30  $\pm$   0.48 \\[2mm]
\multicolumn{4}{c}{\em NGC~5236}\\[1mm]
03 *  &    8.59 $\pm$    0.13  &   $-$0.78  $\pm$   0.28 &   $-$1.59  $\pm$   0.17 \\
\phantom{$^a$}05\footnote{O/H is a lower limit, N/O and S/O are upper limits (see note in Table~\ref{fluxes5236})} \phantom{*}  &    7.78 $\pm$    0.10  &   $-$0.22  $\pm$   0.16 &   $-$1.10  $\pm$   0.12 \\
06 *  &    8.32 $\pm$    0.30  &   $-$0.49  $\pm$   0.64 &   $-$1.24  $\pm$   0.46 \\
09 \phantom{*}  &    7.76 $\pm$    0.19  &   $-$0.16  $\pm$   0.30 &   $-$1.08  $\pm$   0.26 \\
11 *  &    8.94 $\pm$    0.09  &   $-$0.39  $\pm$   0.13 &   $-$1.23  $\pm$   0.12 \\
14 *  &    8.48 $\pm$    0.34  &   $-$0.49  $\pm$   0.64 &   $-$1.43  $\pm$   0.56 \\
16 *  &    8.78 $\pm$    0.12  &   $-$0.68  $\pm$   0.22 &   $-$1.29  $\pm$   0.18 \\